\documentclass[twocolumn]{aastex61}
\pdfoutput=1 
\usepackage{amsmath,amstext}
\usepackage[T1]{fontenc}
\usepackage{apjfonts} 
\usepackage[figure,figure*]{hypcap}

\submitjournal{ApJS}
\received{18 Oct 2017}
\revised{25 January 2018}
\accepted{26 January 2018}
\published{11 May 2018}

\shorttitle{OSSOS: The Complete Data Release}
\shortauthors{Bannister et al.}

\begin{document}

\title{OSSOS: VII. 800+ trans-Neptunian objects --- the complete data release}

\author[0000-0003-3257-4490]{Michele T. Bannister}
\correspondingauthor{Michele T. Bannister}
\email{michele.t.bannister@gmail.com}
\affiliation{Astrophysics Research Centre, School of Mathematics and Physics, Queen's University Belfast, Belfast BT7 1NN, United Kingdom}
\affiliation{Herzberg Astronomy and Astrophysics Research Centre, National Research Council of Canada, 5071 West Saanich Rd, Victoria, British Columbia V9E 2E7, Canada}
\affiliation{Department of Physics and Astronomy, University of Victoria, Elliott Building, 3800 Finnerty Rd, Victoria, BC V8P 5C2, Canada}

\author{Brett J. Gladman}
\affiliation{Department of Physics and Astronomy, University of British Columbia, Vancouver, BC V6T 1Z1, Canada}

\author[0000-0001-7032-5255]{J. J. Kavelaars}
\affiliation{Herzberg Astronomy and Astrophysics Research Centre, National Research Council of Canada, 5071 West Saanich Rd, Victoria, British Columbia V9E 2E7, Canada}
\affiliation{Department of Physics and Astronomy, University of Victoria, Elliott Building, 3800 Finnerty Rd, Victoria, BC V8P 5C2, Canada}

\author[0000-0003-0407-2266]{Jean-Marc Petit}
\affiliation{Institut UTINAM UMR6213, CNRS, Univ. Bourgogne Franche-Comt\'e, OSU Theta F25000 Besan\c{c}on, France}

\author[0000-0001-8736-236X]{Kathryn Volk}
\affiliation{Lunar and Planetary Laboratory, University of Arizona, 1629 E University Blvd, Tucson, AZ 85721, USA}

\author[0000-0001-7244-6069]{Ying-Tung Chen}
\affiliation{Institute of Astronomy and Astrophysics, Academia Sinica; 11F of AS/NTU Astronomy-Mathematics Building, Nr. 1 Roosevelt Rd., Sec. 4, Taipei 10617, Taiwan}

\author[0000-0003-4143-8589]{Mike Alexandersen} 
\affiliation{Institute of Astronomy and Astrophysics, Academia Sinica; 11F of AS/NTU Astronomy-Mathematics Building, Nr. 1 Roosevelt Rd., Sec. 4, Taipei 10617, Taiwan}
\affiliation{Department of Physics and Astronomy, University of British Columbia, Vancouver, BC V6T 1Z1, Canada}

\author{Stephen D. J. Gwyn}
\affiliation{Herzberg Astronomy and Astrophysics Research Centre, National Research Council of Canada, 5071 West Saanich Rd, Victoria, British Columbia V9E 2E7, Canada}

\author{Megan E. Schwamb}
\affiliation{Gemini Observatory, Northern Operations Center, 670 North A'ohoku Place, Hilo, HI 96720, USA}
\affiliation{Institute of Astronomy and Astrophysics, Academia Sinica; 11F of AS/NTU Astronomy-Mathematics Building, Nr. 1 Roosevelt Rd., Sec. 4, Taipei 10617, Taiwan}

\author{Edward Ashton}
\affiliation{Department of Physics and Astronomy, University of British Columbia, Vancouver, BC V6T 1Z1, Canada}

\author{Susan D. Benecchi}
\affiliation{Planetary Science Institute, 1700 East Fort Lowell, Suite 106, Tucson, AZ 85719, USA}

\author{Nahuel Cabral}
\affiliation{Institut UTINAM UMR6213, CNRS, Univ. Bourgogne Franche-Comt\'e, OSU Theta F25000 Besan\c{c}on, France}

\author[0000-0001-9677-1296]{Rebekah I. Dawson}
\affiliation{Department of Astronomy \& Astrophysics, Center for Exoplanets and Habitable Worlds, The Pennsylvania State University, University Park, PA 16802, USA}

\author{Audrey Delsanti}
\affiliation{Aix Marseille Universit\'e, CNRS, LAM (Laboratoire d’Astrophysique de Marseille) UMR 7326, 13388, Marseille, France}

\author[0000-0001-6680-6558]{Wesley C. Fraser} 
\affiliation{Astrophysics Research Centre, School of Mathematics and Physics, Queen's University Belfast, Belfast BT7 1NN, United Kingdom}

\author[0000-0002-5624-1888]{Mikael Granvik}
\affiliation{Department of Physics, P.O. Box 64, 00014 University of Helsinki, Finland}

\author{Sarah Greenstreet}
\affiliation{Las Cumbres Observatory, 6740 Cortona Dr., Suite 102, Goleta, CA 93117, USA}

\author{Aur\'elie Guilbert-Lepoutre}
\affiliation{Institut UTINAM UMR6213, CNRS, Univ. Bourgogne Franche-Comt\'e, OSU Theta F25000 Besan\c{c}on, France}

\author{Wing-Huen Ip}
\affiliation{Institute of Astronomy, National Central University, Taoyuan 32001, Taiwan}
\affiliation{Space Science Institute, Macau University of Science and Technology, Macau}

\author[0000-0002-4385-1169]{Marian Jakubik}
\affiliation{Astronomical Institute, Slovak Academy of Science, 05960 Tatranska Lomnica, Slovakia}

\author{R. Lynne Jones}
\affiliation{University of Washington, Washington, USA}

\author{Nathan A. Kaib}
\affiliation{HL Dodge Department of Physics \& Astronomy, University of Oklahoma, Norman, OK 73019, USA}

\author[0000-0002-1708-4656]{Pedro Lacerda}
\affiliation{Astrophysics Research Centre, School of Mathematics and Physics, Queen's University Belfast, Belfast BT7 1NN, United Kingdom}

\author[0000-0003-2231-3414]{Christa Van Laerhoven}
\affiliation{Department of Physics and Astronomy, University of British Columbia, Vancouver, BC V6T 1Z1, Canada}

\author[0000-0001-5368-386X]{Samantha Lawler}
\affiliation{Herzberg Astronomy and Astrophysics Research Centre, National Research Council of Canada, 5071 West Saanich Rd, Victoria, British Columbia V9E 2E7, Canada}

\author[0000-0003-4077-0985]{Matthew J. Lehner}
\affiliation{Institute of Astronomy and Astrophysics, Academia Sinica; 11F of AS/NTU Astronomy-Mathematics Building, Nr. 1 Roosevelt Rd., Sec. 4, Taipei 10617, Taiwan}
\affiliation{Department of Physics and Astronomy, University of Pennsylvania, 209 S. 33rd St., Philadelphia, PA 19104, USA}
\affiliation{Harvard-Smithsonian Center for Astrophysics, 60 Garden St., Cambridge, MA 02138, USA}

\author[0000-0001-7737-6784]{Hsing~Wen~Lin}
\affiliation{Institute of Astronomy, National Central University, Taoyuan 32001, Taiwan}
\affiliation{Department of Physics, University of Michigan, Ann Arbor, MI 48109, USA}

\author{Patryk Sofia Lykawka}
\affiliation{Astronomy Group, School of Interdisciplinary Social and Human Sciences, Kindai University, Japan}

\author[0000-0001-8617-2425]{Micha\"el Marsset}
\affiliation{Astrophysics Research Centre, School of Mathematics and Physics, Queen's University Belfast, Belfast BT7 1NN, United Kingdom}
 
\author[0000-0001-5061-0462]{Ruth Murray-Clay}
\affiliation{Department of Astronomy and Astrophysics, University of California, Santa Cruz, CA 95064, USA}

\author[0000-0003-4797-5262]{Rosemary E. Pike}
\affiliation{Institute of Astronomy and Astrophysics, Academia Sinica; 11F of AS/NTU Astronomy-Mathematics Building, Nr. 1 Roosevelt Rd., Sec. 4, Taipei 10617, Taiwan}

\author{Philippe Rousselot}
\affiliation{Institut UTINAM UMR6213, CNRS, Univ. Bourgogne Franche-Comt\'e, OSU Theta F25000 Besan\c{c}on, France}

\author[0000-0001-7032-5255]{Cory Shankman}
\affiliation{Department of Physics and Astronomy, University of Victoria, Elliott Building, 3800 Finnerty Rd, Victoria, BC V8P 5C2, Canada}
\affiliation{City of Toronto, Ontario, Canada}

\author{Audrey Thirouin}
\affiliation{Lowell Observatory, 1400 W Mars Hill Rd, Flagstaff, Arizona, 86001, USA}

\author{Pierre Vernazza}
\affiliation{Aix Marseille Universit\'e, CNRS, LAM (Laboratoire d’Astrophysique de Marseille) UMR 7326, 13388, Marseille, France}

\author{Shiang-Yu Wang}
\affiliation{Institute of Astronomy and Astrophysics, Academia Sinica; 11F of AS/NTU Astronomy-Mathematics Building, Nr. 1 Roosevelt Rd., Sec. 4, Taipei 10617, Taiwan}

\begin{abstract}
 
The Outer Solar System Origins Survey (OSSOS), a wide-field imaging program in 2013--2017 with the Canada-France-Hawaii Telescope, surveyed 155 deg$^{2}$ of sky to depths of $m_r = 24.1$--25.2.
We present 838 outer Solar System discoveries that are entirely free of ephemeris bias. This increases the inventory of trans-Neptunian objects (TNOs) with accurately known orbits by nearly 50\%.
Each minor planet has 20--60 \textit{Gaia}/Pan-STARRS-calibrated astrometric measurements made over 2--5 oppositions, which allows accurate classification of their orbits within the trans-Neptunian dynamical populations. 
The populations orbiting in mean-motion resonance with Neptune are key to understanding Neptune's early migration. Our 313 resonant TNOs, including 132 plutinos, triple the available characterized sample and include new occupancy of distant resonances out to semi-major axis $a \sim 130$~au.
OSSOS doubles the known population of the non-resonant Kuiper belt, providing 436 TNOs in this region, all with exceptionally high-quality orbits of $a$ uncertainty $\sigma_{a} \leq 0.1\%$;
they show the belt exists from $a \gtrsim 37$~au, with a lower perihelion bound of $35$~au. We confirm the presence of a concentrated low-inclination $a\simeq 44$ au ``kernel'' population and a dynamically cold population extending beyond the 2:1 resonance.
We finely quantify the survey's observational biases. 
Our survey simulator provides a straightforward way to impose these biases on models of the trans-Neptunian orbit distributions, allowing statistical comparison to the discoveries.
The OSSOS TNOs, unprecedented in their orbital precision for the size of the sample, are ideal for testing concepts of the history of giant planet migration in the Solar System.
\end{abstract}


\section{Introduction}
\label{sec:intro}

We present the full data release of the Outer Solar System Origins Survey (OSSOS), as part of an ensemble of four surveys that together have found 1142 trans-Neptunian objects (TNOs) with well-measured discovery biases. 
We provide a software suite that allows models of the orbital distributions of the trans-Neptunian populations to be ``observed" by OSSOS and the other three well-characterized surveys, imposing their observational biases. The biased models can then be statistically tested against the detected outer Solar System, i.e. the precisely classified TNO discovery sample. Potential dynamical histories of our planetary system can be comprehensively tested.

The dynamical structure of the trans-Neptunian populations are complex and intricate \citep[e.g.][and references therein]{Chiang:2003hb,Gladman2008,Petit:2011p3938,Gladman:2012ed}, a signature of shaping by multiple processes over the last 4.5 Gyr. 
How the TNOs that reside in orbital resonances with Neptune were emplaced is intimately related to how and where the giant planets formed and migrated \citep{Fernandez:1984,kaulanewman92,Malhotra:1993kd,Malhotra:1995dy,thommes02,chiangjordan02,kortenkamp04,tsiganis05,hahnmalhotra05,levison08,Nesvorny:2015et,Kaib:2016tz,pikelawler17}. 
Models of planetary migration result in different distributions of today's resonant objects in (1) the various `types' of libration, (2) the libration amplitude of the resonance angle ($\phi$), (3) their eccentricities \citep[eg.][]{murrayclaychiang05,pikelawler17}, (4) the fraction of resonant TNOs also experiencing Kozai resonance \citep{Lawler:2013hp} and (5) the relative population of different resonances.
As an example, considering the 2:1 mean-motion resonance with Neptune, rapid planetary migration results in fewer objects librating around $\phi \sim90\degr$ than about the other asymmetric island near $\phi \sim 270\degr$, with the fraction depending on the pace and duration of Neptune's early wandering through the Solar System \citep{murrayclaychiang05}. 
The population ratios between particular resonances differ if Neptune's orbital evolution is modelled as a smooth migration \citep{Chiang:2003hb,hahnmalhotra05} or as scattered and damped \citep{levison08}, e.g. between the 2:1, 3:2 and 5:2 resonances \citep{Gladman:2012ed}. 
Three possible mechanisms for resonant capture --- large-scale migration of Neptune \citep[e.g.][]{Malhotra:1995dy,hahnmalhotra05}, chaotic capture of scattered objects \citep{levison08}, and capture of scattered objects through fast secular evolution \citep{Dawson:2012} --- may also result in different libration amplitudes and resonance occupation ratios.

The present set of minor planets suitable for testing models of the Solar System's dynamical history is still small.
Observational surveys of the trans-Neptunian region need to have detailed parametrizations, which make the surveys' discoveries useful for cosmogonic mapping \citep{Kavelaars:ws}. 
The surveys must have precisely measured detection efficiencies, and track essentially all their discoveries in order to avoid ephemeris bias. 
This bias is an insidious loss of unusual orbits from the detected TNO sample.
It happens when the assumed orbit fit to a short ($\lesssim$ month-long) arc of observations is used to predict the future sky location of a TNO for recovery observations made many months later \citep{Jones:2006jl,Kavelaars:ws,Jones:2010bb}. 
TNOs discovered with well-quantified discovery biases are key, as they are appropriate to use to understand the intrinsic population distributions.
As of 2018 January 15, the Minor Planet Center (MPC) database\footnote{\url{https://minorplanetcenter.net/iau/lists/TNOs.html} \url{https://minorplanetcenter.net/iau/lists/Centaurs.html} \url{https://minorplanetcenter.net/iau/lists/NeptuneTrojans.html}} contained 1796 TNOs with orbits known from observations on multiple oppositions (excluding the 167 OSSOS-credited discoveries released to the MPC prior to this date).  
The MPC database is built from the discoveries of more than sixty individual surveys, with a wide variety of inherent and typically unpublished bias characteristics.
Three surveys provide half of the MPC sample.
The two largest past surveys are the Deep Ecliptic Survey (DES 1998--2005; \citealt{Elliot:2005ju}), which provided 478 MPC designations, of which 316 had sufficient observational arc for their orbits to be classified within populations, with 304 secure classifications \citep{Adams:2014hh},
and the Canada-France Ecliptic Plane Survey and its High Ecliptic Latitude extension (CFEPS 2003--2006: \citealt{Jones:2006jl,2009AJ....137.4917K,Petit:2011p3938}, HiLat 2006--2009: \citealt{Petit:2017ju}), which provided 190 secure orbits. 
The Pan-STARRS1 survey (2010--ongoing; \citealt{Chambers.2016}) has provided 370 new objects thus far \citep{Holman:2015,Holman:2017,Weryk:2016}, but these are yet without quantification of their observational biases \citep{Holman:2017}.

We designed the Outer Solar System Origins Survey 
to discover and track many hundreds of distant minor planets, with careful quantification of the bias effects inherent in the observing and analysis of TNOs \citep[see][]{Bannister:2016a}. 
A core theme of OSSOS was to find the fraction of resonant TNOs and map out the filigreed structure of the resonances. 
OSSOS operated as the top-ranked Large Program on the Canada-France-Hawaii Telescope (CFHT) from 2013A through 2016B, with additional observing through 2017.
Our design was based on experience from CFEPS and the survey of \citet[hereafter A16]{Alexandersen:2016ki}. 
We surveyed 155.3 deg$^2$ of sky near the invariable plane, at a range of heliocentric longitudes, to moving-target limiting magnitudes ($m_r$) ranging from $m_r = 24.1$ to 25.2.
Three-quarters of the survey goes deeper than the forthcoming Large Synoptic Survey Telescope will reach in single observations \citep{Jones:2016}. 

Our full survey data release focusses on the products from our five years of observations, 2013 January through 2017 December. The calibration and analysis of the images are given in \S~\ref{sec:observations}. The entire survey's discoveries of TNOs are detailed in \S~\ref{sec:discoveries}.
Of our TNOs, 840 were bright enough to have detailed bias-characterization analysis. 
Of these, 838 have received tracking observations that ensure that the objects can never be lost, while merely 2 close Centaurs were lost. 
No previous survey has ever achieved such a high rate of success in tracking TNOs to good orbits. 
Our established survey simulator is expanded and improved from that in \citet{Petit:2011p3938}. 
It permits assessment of models of the trans-Neptunian populations against a calibrated and carefully tracked set of TNO detections. We provide simulator files detailing the observations and the characterized detections (1142 in total) from an ensemble of four bias-characterized surveys with CFHT: CFEPS, HiLat, A16 and OSSOS (\S~\ref{sec:framework}).
This framework is ideal for investigating the processes at work during the formation and early evolution of the Solar System.
We conclude by highlighting the orbital properties of our discoveries, with comparison to the MPC inventory (\S~\ref{sec:populations}).

\section{Data acquisition and calibration}
\label{sec:observations}

\begin{figure*}
\plotone{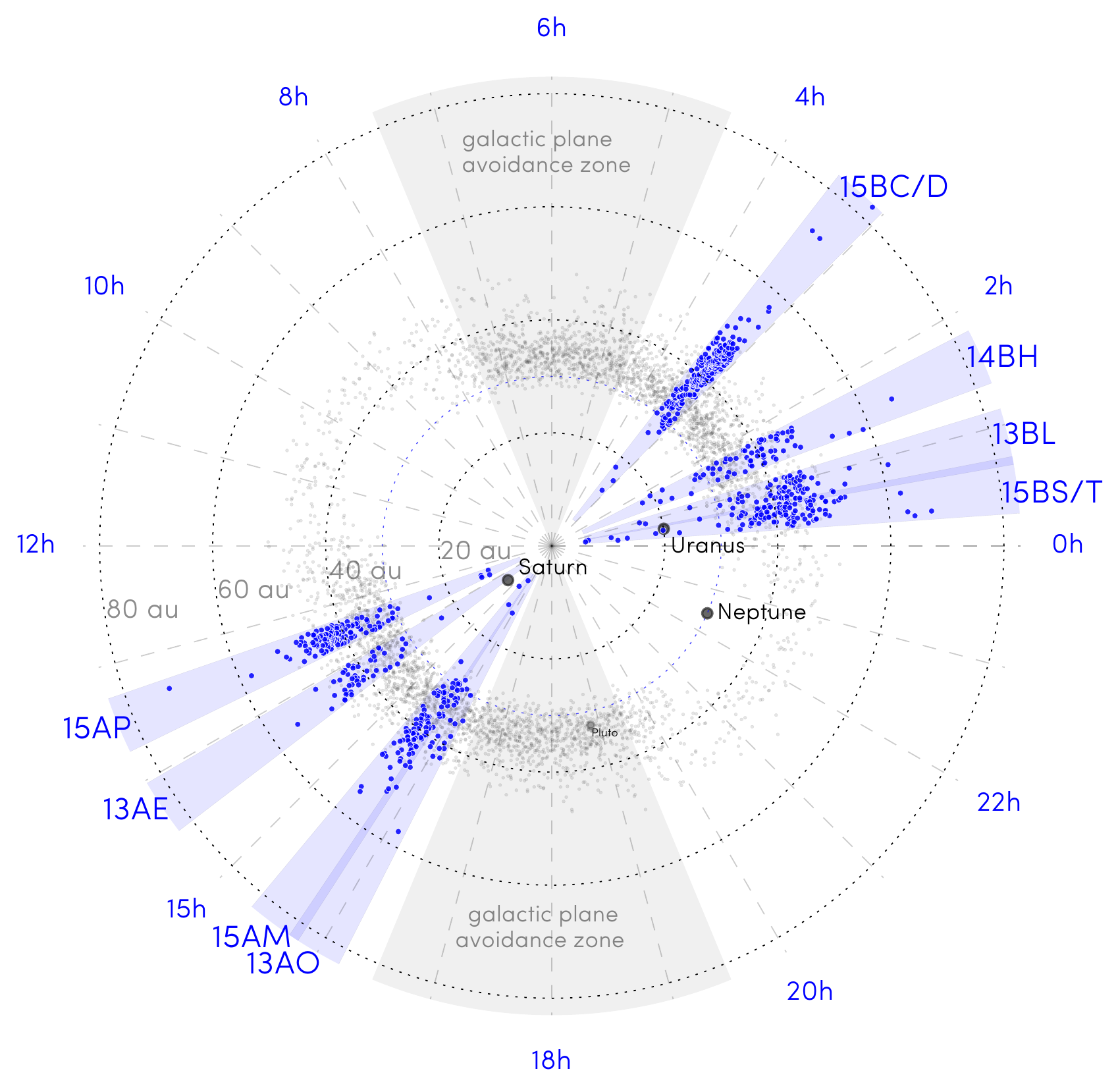}
\caption{The spatial relationship of the regions of sky targeted by OSSOS to the geometry of the outer Solar System. The eight sky blocks are indicated by blue wedges, flattened from their low 0--10\degr inclinations into the plane (the on-sky projection is shown in Fig.~\ref{fig:blocks}). The blocks were placed to avoid the dense star fields of the galaxy (schematically indicated by grey shading). Grey dots show the predicted position density of the observable fraction ($m_r < 24.7$) of objects in the 3:2 resonance with Neptune, as modelled by \citet{Gladman:2012ed}. Blue dots are the 840 characterized OSSOS discoveries (Table~\ref{tab:discoveries}), which were found at heliocentric distances between 6 and 83~au. The sensitivity of OSSOS to distant moving objects extends beyond the figure boundaries to $\sim 100$--$130$~au and is discussed in \S~\ref{sec:analysis}.}
\label{fig:pointings}
\end{figure*}

All observations were acquired with the 0.\arcsec184/px MegaCam imager \citep{2003SPIE.4841...72B} of CFHT on Maunakea, Hawai'i.
\citet[][\S~2]{Bannister:2016a} details the survey design; here we provide a brief summary. 
OSSOS surveyed the distant Solar System objects present in eight regions of sky (``blocks", each $\sim20$ deg$^{2}$).
The spatial relationship of these pointings to one of the target resonant TNO populations, the plutinos (3:2 mean-motion resonance with Neptune), is shown in Fig.~\ref{fig:pointings}. The full geometry of the survey is in Table~\ref{tab:pointings}.

\begin{figure*}
\gridline{\fig{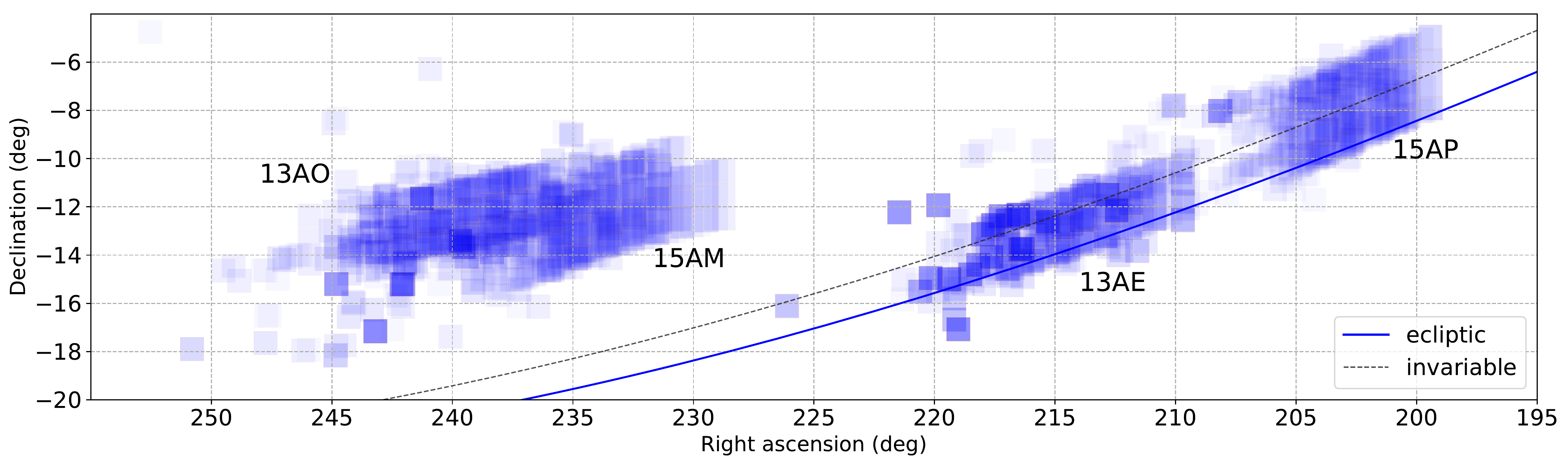}{\textwidth}{(a) OSSOS survey blocks with April--May oppositions.
\label{fig:Asemester}}
}
\gridline{\fig{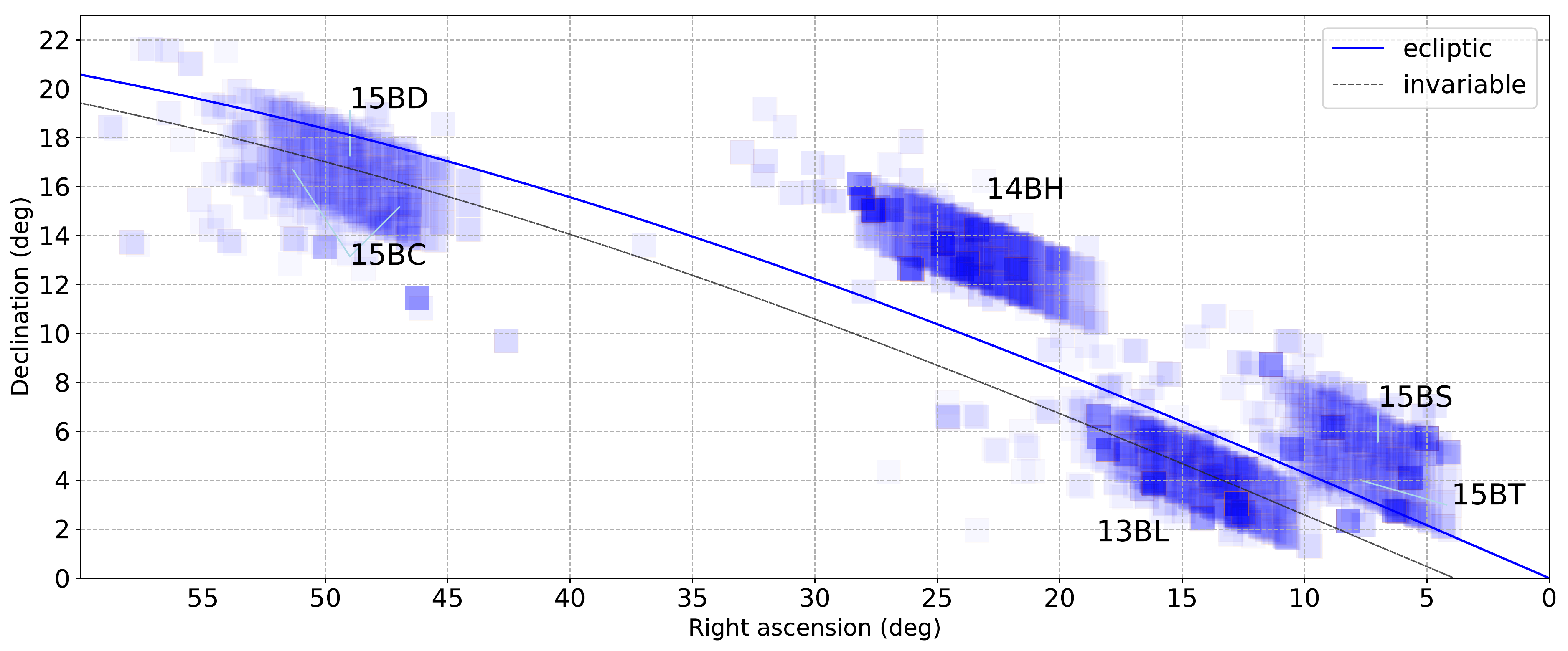}{\textwidth}{(b) OSSOS survey blocks with September--November oppositions. Several blocks are described for TNO detection efficiency characterization in two parts, as their discovery observations had different limiting magnitudes. 15BS is the upper and 15BT the lower half of a single region (\S~\ref{sec:15BS}). The two columns of 15BC bracket the deeper central region of 15BD (\S~\ref{sec:15BD}) Exact boundaries at discovery are given in Table~\ref{tab:pointings} and in the survey simulator (\S~\ref{sec:framework}). 
\label{fig:Bsemester}}
}
\caption{Sky locations of the OSSOS survey blocks, showing all CFHT MegaCam imaging 2013--2017 (blue shading).}
\label{fig:blocks}
\end{figure*}

The OSSOS blocks were placed on the sky at a range of low latitudes chosen relative to the invariable plane of the Solar System \citep{Souami:2012fc}, rather than to the ecliptic (Fig.~\ref{fig:blocks}).
The choice of reference plane when siting survey regions is important for low-latitude surveys. The inclination distributions of the dynamical populations of TNOs are well described by overlapping Gaussians of various widths \citep{Brown:2001p3803}: the narrowest, that of the cold classical Kuiper belt, has a mere $\sigma \sim2\degr$ \citep{Gulbis:2010fw,Petit:2011p3938}. 
The mean plane of the Kuiper belt is certainly not the ecliptic, nor is it precisely the invariable plane, and it is not consistently flat with increasing semi-major axis \citep{Brown:2001p3803,Chiang:2008dz,Volk:2017}. However, the invariable plane provides a reasonable proxy for the purpose of large-area survey design.
As the spacing on the sky between the ecliptic and invariable planes varies with longitude by up to $\sim4\degr$, proximity to a given plane will affect the detection rates of the cold classical population.
Each OSSOS block was a rectangular grid, arranged to best tessellate MegaCam's $\sim 0.9$ deg$^2$ field of view parallel to the invariable plane and reduce shear loss of TNOs (Figure~\ref{fig:blocks}).

Discovery observations were made when each block of sky came to opposition, and they set the limiting magnitudes of the survey. 
OSSOS observed in the $r$-band ($\lambda \sim 640$ nm) MegaCam filters $R.9601$ and $R.9602$, which approximate the $r$-band SDSS response \citep{Fukugita:1996}, and in a ``$w$" wide-band filter, $GRI.MP9605$ --- all of which are specified in \S~\ref{sec:photcalib} and in Fig.~\ref{fig:ossosfilters}.
Each of the \textit{fields} in a block was observed with a triplet of exposures in an $r$-band MegaCam filter, spaced over two hours. 
The image quality of this discovery triplet\footnote{The CFHT image IDs of the images used in the discovery triplet for each block are provided in the Supplemental Materials. See Footnote 3 for image access details.} was crucial: the depth reached in the discovery images set the detectable flux limit of TNO discoveries for OSSOS, which in combination with the tracking efficiency, determined each block's \textit{characterization limits}. 
These derived brightness- and sky motion rate-dependent limiting magnitudes for OSSOS are given in Table~\ref{tab:characterization}.

OSSOS was designed to provide the necessary orbital accuracy of its TNOs via (a) a dense cadence, with observations in every available dark run over two consecutive years, and (b) through the elimination of systematic errors in astrometry catalogues (discussed in \S~\ref{sec:calibrations}).
As in the A16 survey, the observations in the first year targeted only each block's large grid of pointings. 
We offset the sky position of the discovery fields throughout the observing semester, at a rate consistent with the mean rate of motion for TNOs expected to be detectable in the field. 
This observing technique maximised observations of the TNOs that would be present in the block at opposition, and ensured that the TNOs were imaged at least several times each lunation. 
Our avoidance of estimating follow-up ephemerides from short observational arcs removed ephemeris bias.

The high astrometric precision and numerous observations in the first year ensured that ephemeris predictions in the second year generally had uncertainties smaller than an arc\-minute, allowing observations to use targeted pointings instead of a grid.
The pointings were designed to recover all objects as they slowly sheared further apart; the TNOs were still clustered close enough on the sky to frequently observe multiple TNOs in each pointing.
Tracking observations were at first made in an $r$-band filter, then, after 2015 August, instead in the ``$w$'' $GRI.MP9605$ wide-band filter (\S~\ref{sec:calibrations}). 
The additional observations refined the TNOs' orbital parameters, until they were better than a quality threshold: we required a fractional semi-major axis uncertainty $\sigma_{a} <0.1\%$ (see \S~\ref{sec:orbitquality}).

Considering purely the overall dataset of images, the dense OSSOS cadence provided 20-60 epochs across 2-5 years, each to an $r$- or $w$-band $\sim 3\sigma$ depth of magnitudes 24.1-25.3, across a substantial region of sky: $\sim170$ deg$^2$ in the vicinity of the ecliptic. The density of visits is indicated by shading in Fig.~\ref{fig:blocks}. More than eight thousand images were acquired, making OSSOS one of the most data-rich surveys yet made with CFHT. The depth of the combined imaging if stacked would approach $m_r \sim 26.3$ across large areas. 
All the calibrated survey images are available\footnote{\doi{10.11570/18.0001}} from the Canadian Astronomy Data Centre (CADC).

\citet{Bannister:2016a} describes the observations of the first two blocks (13AE, 13AO) in 2013--2015.
We now describe the observations of the remaining six blocks, which had their discovery observations in the semesters 2013B, 2014B, 2015A and 2015B (\S~\ref{sec:13BL}--\ref{sec:15BD}). 
We also detail observations made for OSSOS with CFHT in 2014--2017 outside of the Large Program (\S~\ref{sec:additional-obs}).
We then discuss our astrometric and photometric calibration of the images (\S~\ref{sec:calibrations}), and quantify our analysis pipeline's TNO detection efficiency (\S~\ref{sec:analysis}).

\begin{deluxetable*}{llllllcccr}
\tabletypesize{\footnotesize}
\tablecolumns{10}
\tablecaption{Areas of the sky observed for discovery by OSSOS \label{tab:pointings}}
\tablehead{\colhead{Block} \vspace{-0.2cm} & \colhead{RA} & \colhead{Dec} & \colhead{Epoch} & \colhead{Field} & \colhead{Area} & \colhead{Filling} & \colhead{Filter} & \colhead{$m_{limit}$} & \colhead{TNOs} \\
\colhead{ } & \colhead{(hr)} & \colhead{(\arcdeg)} & \colhead{(MJD)} & \colhead{layout} & \colhead{(deg$^2$)} & \colhead{factor} & \colhead{ } & \colhead{(3\arcsec/hr)} & \colhead{detected}
}
\startdata
15BS & 00:30:08.35 & +06:00:09.5 & 57274.42965 & 2 x 5 & 10.827 & 0.9223 & R.MP9602 & 25.12 & 67 \\
15BT & 00:35:08.35 & +04:02:04.5 & 57273.42965 & 2 x 5 & 10.827 & 0.9223 & R.MP9602 & 24.93 & 54 \\
13BL & 00:52:55.81 & +03:43:49.1 & 56596.22735 & 3 x 7 (-1) & 20.000 & 0.9151 & R.MP9601 & 24.42 & 83\\
14BH & 01:35:14.39 & +13:28:25.3 & 56952.27017 & 3 x 7 & 21.000 & 0.9103 & R.MP9601 & 24.67 & 67\\
15BC & 03:06:18.32 & +15:31:56.3 & 57332.33884 & 1 x 4 & 4.3309 & 0.9215 & R.MP9602 & 24.78 & \\
15BD & 03:12:58.36 & +16:16:06.4 & 57333.35377 & 2 x 4 & 8.6618 & 0.9211 & R.MP9602 & 25.15 & 146 \\
15BC & 03:22:09.15 & +17:15:44.0 & 57332.33884 & 2 x 4 & 8.6618 & 0.9215 & R.MP9602 & 24.78 & 104 \\
15AP & 13:30:22.11 & -07:47:23.0 & 57125.36971 & 4 x 5 & 21.654 & 0.9186 & R.MP9602 & 24.80 & 147 \\
13AE & 14:15:28.89 & -12:32:28.5 & 56391.36686 & 3 x 7 & 21.000 & 0.9079 & R.MP9601 & 24.09 & 49 \\
15AM & 15:34:41.30 & -12:08:36.0 & 57163.31831 & 4 x 5 & 21.654 & 0.9211 & R.MP9602 & 24.87 & 87 \\
13AO & 15:58:01.35 & -12:19:54.2 & 56420.45956 & 3 x 7 & 21.000 & 0.9055 & R.MP9601 & 24.40 & 36\\
\enddata
\tablecomments{The filling factor correction is discussed in \S~\ref{sec:analysis}: it incorporates the true pixel area, the small overlap area due to the new shape of the CCD in discovery blocks from 2015, and the few incompletely searched chips. It is used in the survey simulator (\S~\ref{sec:framework}) when testing the visibility of model objects by their location on the sky.
The survey simulator uses a single date for each block, as that is statistically equivalent: the simulator produces a statistical ensemble that is representative of the detections, and the approximation provides computational efficiency.
Note that 15BC is in two parts (\S~\ref{sec:15BD}) and thus appears twice in this table; however, its detections are given once as a total.
Limiting magnitude is given for sources with a sky motion rate of 3 \arcsec/hr: for comprehensive details across the full range of motion rates, see \S~\ref{sec:analysis} and Table~\ref{tab:characterization}.
TNO detections are discussed in \S~\ref{sec:discoveries}.
(This table is incorporated in machine-readable form with fully specified polygons inside the survey simulator, linked in \S~\ref{sec:framework}). 
}
\end{deluxetable*}

\subsection{Block 13BL observations}
\label{sec:13BL}

The 13BL block was a 3 x 7 grid of MegaCam fields that overlay the invariable plane, centred at 0\fh55\arcmin, 4\arcdeg00\arcmin\ (Fig.~\ref{fig:Bsemester}).
The 13BL discovery observations were made in median image quality (IQ) of $0.75$\arcsec\ with the $R$.MP9601 filter.
Half the block was acquired on 2013 September 29, and the other half acquired on 2013 October 31.
A complete triplet sequence was acquired on 20 MegaCam fields (one fewer than in the 21-field block design).
Uranus was about a degree away during discovery acquisition, but did not contribute any scattered light.
Tracking observations within the discovery lunation and in the lunations either side were smoothly acquired.
Pointed tracking observations were made from 2014 July through 2015 January, all with the $R.MP9601$ filter.

\subsection{Block 14BH observations}
\label{sec:14BH}

The 14BH block was a 3 x 7 grid of pointings, 2--5\arcdeg off the invariant plane, centred at 1\fh29\arcmin, 12\arcdeg58\arcmin\ (Fig.~\ref{fig:Bsemester}).
Poor weather during the 2013 October opposition prevented observation of a valid discovery triplet for the whole H block. 
Only six fields received a survey-quality $m_r \sim 24.5$ depth triplet on 2013 November 1. 
Fortunately, the multi-year nature of the Large Program allowed the discovery observations for H block to be deferred to the next year's opposition.
Also, the CFHT mirror was re-aluminised in 2014 July, a process that increases optical throughput.
The full 21 fields of 14BH received discovery triplet observations on a single night, 2014 October 22, in the $R.MP9601$ filter.
All discoveries at Kuiper belt distances found in the six-field 2013 triplet were re-discovered in the 2014 triplet.
The capriciousness of weather meant the designed OSSOS survey cadence was modified for this block: there were two semesters of only tracking the 14BH grid, with the discovery triplet observed in the second. 
The cadence and strategy remained successful under this strain.
During analysis, the density of the 2013 observations permitted immediate linking out of each TNO's orbit from the arcs in the 2014 discovery semester to ``precovery" arcs throughout 2013B. 
Despite the unorthodox observing, only minimal clean-up observations with pointed ephemerides were needed in 2014 December and 2015 January. These secured the orbits of a small fraction of $<30$ au discoveries, which had moved to areas of sky outside the grid of the block that was observed in 2013--14.

\subsection{Block 15AP observations}
\label{sec:15AP}

The 15AP block was a 4 x 5 grid of pointings, spanning the invariant plane and centred at 13\fh30\arcmin, -~7\arcdeg45\arcmin\ (Figure~\ref{fig:Asemester}). Its positioning, lower in R.A. than 13AE, centres it more on the invariant plane, while avoiding potential scattered light from $\alpha$ Vir.
15AP was acquired for discovery on 2015 April 12 in 0.6\arcsec seeing in the $R$.MP9602 filter (discussed below). 

2015 saw the introduction of a new set of filters for CFHT's MegaPrime. 
The new filters allowed use of four chips that have been on MegaCam since its commissioning, but were vignetted out. 
This changed the MegaCam focal plane from 36 chips to 40 chips, which required rearranging the OSSOS grid to a 4 x 5 tessellation.
All 2015A observations, including the discovery triplets, were acquired in MegaCam's new $R$.MP9602 filter. 
\S~\ref{sec:photcalib} discusses how $R$.MP9602 has a very similar bandpass, but 0.13 mag higher throughput than the $R$.MP9601 used in previous years (Fig.~\ref{fig:ossosfilters}).
Use of the $R$.MP9602 maintained consistency in the discovery bandpass across the whole of OSSOS. 
Section~\ref{sec:framework} discusses how TNO colours can thus be modelled.

The new wide-band filter for MegaCam, GRI.MP9605, spans 4000 to 8200~\AA\ and is very flat in transmission. 
We discuss its characteristics in \S~\ref{sec:calibrations} (Fig.~\ref{fig:ossosfilters}).
Observations in GRI.MP9605 are referred to as $w$ in our observations reported to the Minor Planet Center.
We used the GRI.MP9605 filter exclusively for all our tracking observations after it became available in 2015 August: it added an extra $\sim 0.7$ magnitudes of depth (\S~\ref{sec:photcalib}), ideal for recovery of fainter TNOs near the characterisation limit.

Good weather in the 15A semester allowed pointed recoveries in GRI.MP9605 on the TNOs discovered in 15AP to proceed without incident from January through late July 2016.
Two observations, spaced by a week to ten days, were made each lunation.

\subsection{Block 15AM observations}
\label{sec:15AM}

The off-plane 15AM block was a 4 x 5 grid of pointings nestled next to the earlier 13AO block \citep{Bannister:2016a} at a slightly lower ecliptic latitude, centred at 15\fh30\arcmin, -~12\arcdeg20\arcmin\ (Figure~\ref{fig:Asemester}).
The multi-year nature of our Large Program permitted advance observations: we predicted the 2014 location of model Kuiper belt objects that would be within the desired region of sky at opposition in 2015. 
In 2014 we imaged 21 deg$^2$ (arranged in the 36-chip 3 x 7 grid layout) with a triplet in $R$.MP9601, with half on 2014 May 29 in 0.5--0.7\arcsec seeing and the remainder on 2014 June 1 in similar image quality.
In 2015, a successful discovery triplet in the new $R$.MP9602 filter (see Section~\ref{sec:15AP}) saw 0.5\arcsec IQ on each half, observed respectively on 2015 May 24 and 25.
Despite the change in the grid layout (see Section~\ref{sec:15AP}), and the comparatively higher dispersal rate on the sky of TNOs on more inclined and eccentric orbits, the 2014 observations were able to provide precovery for more than half of the 15AM discoveries.
Tracking throughout 16A in the new GRI.MP9605 filter proceeded smoothly (see Section~\ref{sec:15AP}).

\subsection{Block 15BS-15BT observations}
\label{sec:15BS}

The 15B observations provide the deepest areas of sky observed by OSSOS, with discovery observations in MegaCam's new filters $R$.MP9602 and tracking observations in GRI.MP9605 (see Section~\ref{sec:15AP}), in a semester featuring exceptionally good weather.

The 15BS block was a 4 x 5 grid of pointings, placed above the invariant plane to sample inclinations midway between those observed by 13BL and 14BH, and centred at 0\fh30\arcmin, +05\degr00\arcmin\ (Figure~\ref{fig:Bsemester}). 
Its discovery triplet acquisition in $R$.MP9602 in 2015B was in two halves, split parallel to the plane. 
Both were observed in excellent seeing. 
However, the half observed on 2015 September 9 had better IQ than that on 2015 September 8, and the half acquired on September 8 had a small pointing shift error for one image of the triplet. 
For characterization we thus regard it as two adjacent blocks of equal size: 15BS, the southern half, shallower and closer to the invariant plane, and 15BT, deeper and higher-latitude, as shown in Figure~\ref{fig:Bsemester}. 

The tracking acquisitions in GRI.MP9605 in 2015B proceeded smoothly.
However, 2016B saw several dark runs be near total losses due to weather. A quarter of the 2015B discoveries from 15BS/15BT received insufficient tracking for high-precision orbit determination, and were targeted for additional 2017 observation.  

\subsection{Block 15BC-15BD observations}
\label{sec:15BD}

The 15BD block was a 4 x 5 grid of pointings, centred on the invariant plane at 03\fh15\arcmin, +16\degr30\arcmin\ (Figure~\ref{fig:Bsemester}).
As with the 15AM block (Section~\ref{sec:15AM}), advance planning allowed acquisition of precovery imaging in 2014. 
This included acquisition on 2014 November 17 of a lengthened, 400-s exposure triplet in $R$.MP9601 on 11 deg$^2$ of the block.
Designed to be the deepest block OSSOS would acquire, the 2015 discovery triplet's exposure times were lengthened to 400 s in $R$.MP9602. We redesigned the acquisition cadence to ensure that a triplet spanning two hours could still be successfully acquired on a large enough area of sky.
Notable variation in IQ occurred between the two nights of observation: 0.6\arcsec\ IQ on 2015 November 6, and 0.45\arcsec\ IQ on 2015 November 7. The characterization on this block is therefore also split into two, forming the adjacent 15BC and 15BD areas (Figure~\ref{fig:Bsemester}).
Both have the same latitude coverage. The shallower 15BC is in two parts: a 1 x 4 deg$^2$ column and a 2 x 4 deg$^2$ area, as shown in Figure~\ref{fig:Bsemester}. These bracket the contiguous 2 x 4 deg$^2$ area of 15BD, which is the deepest area imaged by OSSOS.

Similar to that for block 15BS (see \S~\ref{sec:15BS}) the tracking observations in GRI.MP9605 in 15B were well sampled. 15BD was buffered against the sparse observing that occurred in 2016B (see \S~\ref{sec:15BS}) by the partial-coverage observations in 2014. Exposure times for the GRI.MP9605 tracking observations were increased to 450 s to retain the deep discoveries.

\subsection{Additional observations}
\label{sec:additional-obs}

After two years of intensive observation, a small fraction of TNOs either had orbits still classifying as insecure, or were securely resonant but had large libration amplitude uncertainties. These were targeted with a few extra observations in GRI.MP9605: at minimum twice, and in some cases 3 or 4 epochs, at times far from opposition that provided the greatest parallax. 
In 2017, these observations were provided via a Director's Discretionary (17A) and regular-time (17B) proposal.
The CFHT mirror was re-aluminized in 2017 August, again producing slightly deeper images.

Additional observations came from a Director's Discretionary program with CFHT MegaCam in 2014--2017, which simultaneously observed a magnitude-limited $m_r < 23.6$ sample of OSSOS TNOs together with Gemini North for \textit{Colours of OSSOS} (Col-OSSOS; e.g. \citet{Fraser:2015DPS}).
The program observed a uniform \textit{r-u-r} filter sequence, with the two 300 s $r$-band images separated by 1.5 hours. 
The $r$ filter was $R$.MP9601 in 2014B-2015A, and $R$.MP9602 after August 2015. 
The Col-OSSOS imaging covered most of the area of 13AE, 13AO, 13BL, 14BH, 15BS and 15BT \citep{Bannister:2017inprep}. 
Its $r$-band observations are folded into the OSSOS analysis.

The targeted imaging and that for Col-OSSOS produced substantial amounts of serendipitous observing of the TNOs nearby on the sky, due to the large MegaCam field of view.
The orbits (\S~\ref{sec:discoveries}) in this paper thus update those given in \citet{Bannister:2016a} for the discoveries from blocks 13AE and 13AO.

\subsection{Astrometric and photometric calibration}
\label{sec:calibrations}

OSSOS required high-precision, coherent astrometric calibration to produce TNO orbits of the desired quality (\S~\ref{sec:orbitquality}).
During the February 2013--October 2016 tracking of each TNO, we acquired observations and then re-classified the newly improved orbit to determine if additional observations were required to secure the orbit. The image calibration during this time used an iterative approach with the 2MASS and UCAC4 catalogues, described in \citet[][\S~3]{Bannister:2016a}.
Since the Gaia Data Release 1 was published in late 2016, all OSSOS data has been astrometrically calibrated using Gaia.  
At the end of the Large Program observations, all earlier OSSOS images were recalibrated using Gaia.
The orbits that we report here (\S~\ref{sec:discoveries}) and to the Minor Planet Center are based on this calibration.

\subsubsection{Astrometric calibration}
\label{sec:astromcalib}

The \textit{Gaia} Data Release 1 (Gaia-DR1; 
\citet{GaiaDR1:2016}) was used for the principal astrometric reference catalog. The Gaia-DR1 catalog extends to a magnitude of $g \sim 20$. The astrometric uncertainties are typically 10 milliarcseconds (mas) and do not include estimates of proper motion. However, since OSSOS observed from 2013 to 2017, no observation is more than 2 years from the Gaia-DR1 epoch of 2015.0, so proper motion is unlikely to cause significant systematic effects. The majority of the OSSOS images were calibrated directly with Gaia-DR1.

For 201 of the more than eight thousand OSSOS images, the Gaia catalog was insufficiently dense to compute an accurate plate solution. In these cases, we used the Pan-STARRS Data Release 1 catalog (PS-DR1; \citet{PanStarrsDR1:2016}). PS-DR1 extends 2-3 magnitudes deeper than Gaia, with a corresponding increase in source density. It was calibrated with the 2MASS catalogue \citep{Skrutskie:2006hl}, leading to significant zonal errors. Just before the PS-DR1 release the coordinates were corrected. The quality of this correction is generally good: for most fields the typical residuals between PS-DR1 and Gaia-DR1 are on the order of 10 mas. However, there exist several patches of sky in PS-DR1 where there is a considerable offset between PS-DR1 and Gaia. For each of the 201 OSSOS images calibrated with PS-DR1, we examined the residuals between PS-DR1 and Gaia-DR1.  Only two images had an unacceptable match, and were instead calibrated with 2MASS; their astrometric accuracy is significantly worse.

To measure the astrometric errors, the positions of the sources in every overlapping pair of images were compared. The residuals give an indication of the astrometric errors. The results are 29 mas for R.MP9601, 21 mas for R.MP9602 and 49 mas for GRI.MP9605. The difference in accuracy between the two $r$ filters can be explained by two factors. The installation of vents in the dome of CFHT in late 2013 \citep{Bauman:2014SPIE} improved the median seeing in $r$ by $\sim 0.1\arcsec$ \citep{Devost:16} relative to earlier CFHT imaging \citep{Salmon:2009}. Additionally, the new R.MP9602 filter has better throughput (\S~\ref{sec:photcalib}), resulting in deeper images. The comparatively worse residuals for the GRI.MP9605 wide filter are due to differential chromatic refraction (DCR). 

\subsubsection{The effects of differential chromatic refraction (DCR)}
\label{sec:dcr}

The effects of DCR are about 10 times as great on the 4000~\AA-wide GRI.MP9605 as on either of the two $r$-filters. 
However, while DCR affected the positions of individual stars by tens of milli-arcseconds, the net effect on the plate solution was virtually nil. 
On average the plate solution was shifted by 1-2 mas, far smaller than the above-noted random errors. 
The effects of DCR could only be detected due to the high precision of Gaia-DR1. 
The earlier calibration done in \citet{Bannister:2016a} was with respect to 2MASS and UCAC4 \citep{Zacharias:2013cf}, and the random errors in 2MASS overwhelmed the effects of DCR. 

We investigated if it would be possible to roughly determine the colour of our TNOs from their astrometry, as DCR will systematically shift the measured positions of the TNOs with respect to their true positions.
The amplitude of the positional shift from DCR depends on airmass. 
The direction depends on color: towards the zenith (along the complement of the parallactic angle) if bluer than the average star, towards the horizon if redder. 
The amplitude of the shift as a function of zenith angle, $z$, can be computed theoretically using an atmospheric model \citep{AllenAtmos} if one has full knowledge of the throughput of the telescope, camera and filter system, as well as the spectral energy distribution of each observed source. 
The atmospheric model would need to incorporate the hourly weather conditions and the pointing azimuth and altitude to compute the temperature/pressure profile as a function
of altitude, along the line of sight.
We chose instead to measure DCR empirically, in terms of arcseconds of shift per magnitude of $g-i$ relative to the mean $g-i$ color of stars.  
DCR increases linearly with $\tan(z)$. 
We found the slope for the $GRI.MP9605$ filter to be 75 mas/magnitude per unit $\tan(z)$. 
The amplitude of the slope is 7 mas/magnitude per unit $\tan(z)$ for the two $r$-band filters. 
The bulk of our observations were made at an airmass $<1.4$ (or equivalently $\tan(z)<1$). 
The known range of colors of TNOs is smaller than the range of colors of stars: about $\pm$0.5 magnitudes centred on $g-i=1.2$ \citep{Sheppard2012,WongandBrown2016}, which turns out to be also roughly the mean color of stars. 
The shift in sky position for a TNO with a colour towards the edges of the color distribution is typically 30-40 mas, while TNOs with more median colours will have a smaller shift. 
Thus, the effect of DCR on the OSSOS measurements is dwarfed by the median 124 mas centroid uncertainties of each TNO observation, and our astrometry cannot be used to infer the colour of the TNOs. 

\subsubsection{Photometric calibration}
\label{sec:photcalib}

\begin{figure}
\plotone{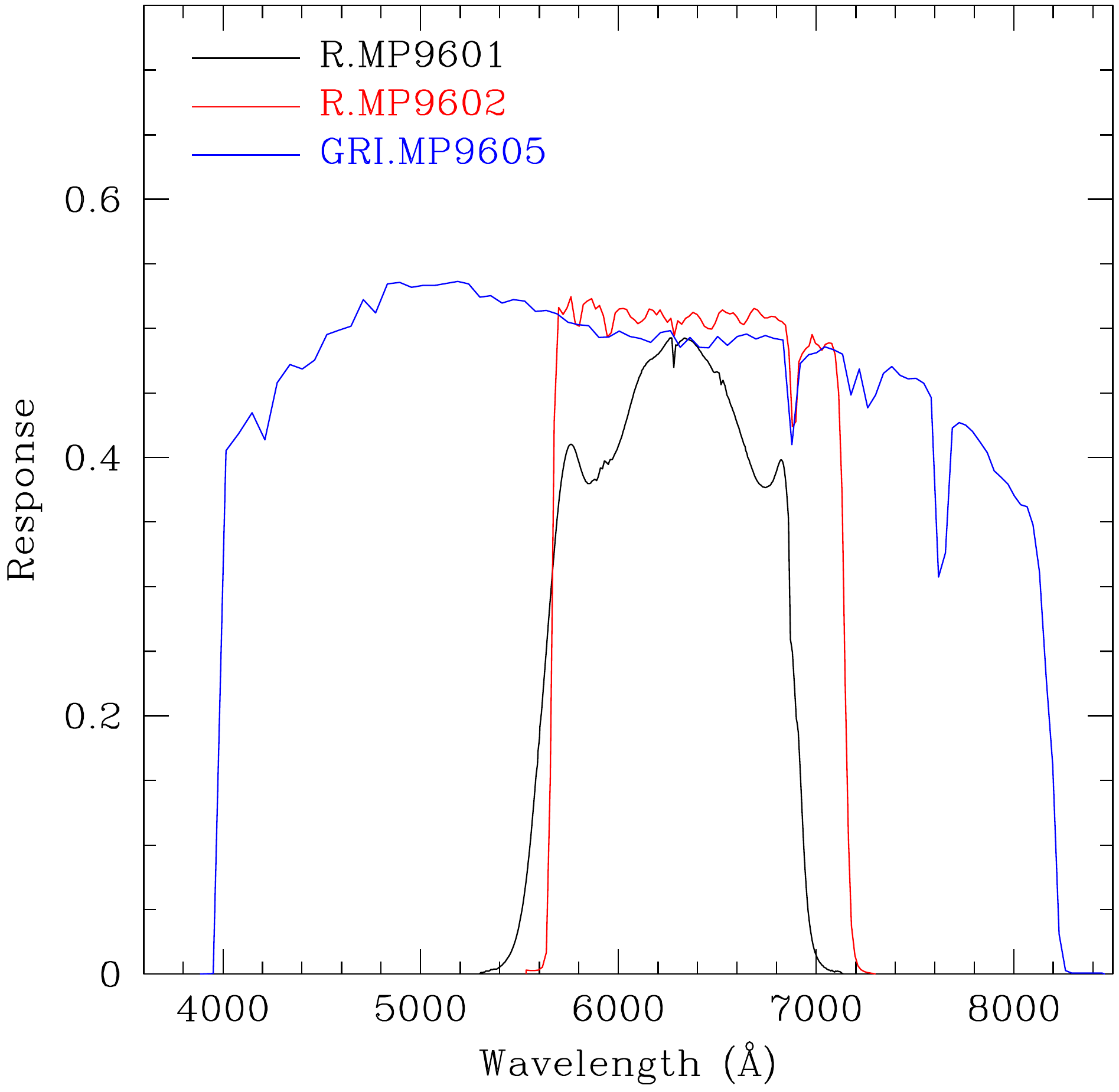}
\caption{Filter responses for the CFHT MegaCam filters used in OSSOS. Discovery imaging was made exclusively in the two very similar $r$ filters (Table~\ref{tab:pointings}). Most tracking images were made with the wide GRI.MP9605 filter, which we refer to in our astrometric measurement lines as $w$. The detailed data for the filter response curves is available at \url{http://www.cadc-ccda.hia-iha.nrc-cnrc.gc.ca/en/megapipe/docs/filt.html}
\label{fig:ossosfilters}}
\end{figure}

The filter responses for the three OSSOS filters are shown in Figure \ref{fig:ossosfilters}. 
These filter responses include the full system of the telescope and camera (including mirror, lenses and the quantum efficiency of the detector), 1.25 airmasses of atmosphere as well as the transmission of the filters themselves. The newer r-band filter (R.MP9602) is slightly wider than the older r-band filter (R.MP9601) and has higher transparency over a wider wavelength range. Images taken in R.MP9602 are on average 0.13 magnitudes deeper than those taken with R.MP9601. Images taken in the wide filter (GRI.MP9605) are on average 0.71 magnitudes deeper than those taken in R.MP9601.

Previously, the photometric calibration was based on the Sloan Digital Sky Survey (SDSS; \citet{Fukugita:1996,Ahn:2014fa}). As described in \citet[][\S~3]{Bannister:2016a}, images taken on the footprint of the SDSS were directly calibrated using the SDSS catalogue converted into the MegaCam system. For images off the SDSS footprint, the procedure was more intricate. First, we computed the zero-point variation across the focal plane of MegaCam. Then, for each photometric night, all images on the SDSS were used to compute a nightly zero-point. Images taken on photometric nights were calibrated using this zero-point. Images taken on non-photometric nights were calibrated by transferring the zero-point from images taken on photometric nights using a bootstrapping system.

With the release of PS-DR1, this system was replaced with a much simpler system. Each image was calibrated using the PS-DR1 photometric catalogue. The Pan-STARRS photo\-metry was converted to the MegaCam photometric system:

\begin{align}
R.MP9601   &=r+0.002 -0.017x+0.0055x^2-0.00069x^3 \\
R.MP9602   &=r+0.003 -0.050x+0.0125x^2-0.00699x^3 \\
GRI.MP9605 &=r+0.005 +0.244x-0.0692x^2-0.00140x^3 
\end{align}

where $r$ refers to $r_{\rm Pan-STARRS}$ and $x=g-i$ in the Pan-STARRS system. This third order polynomial expression of color terms in $g-i$ follows the convention in \citet{2016ApJ...822...66F}. The transformation for GRI.MP9605 is only valid for $g-i < 1.4$. Sources redward of this limit were not used, and only stars (not galaxies) were used. The transformed Pan-STARRS stellar photometry was used to calibrate each CCD of each image separately. Occasionally, the number of stars on a CCD was insufficient to robustly calibrate that CCD.  In these cases, the zero-point was determined for the focal plane as a whole, and that zero-point was applied to the individual CCD. Magnitudes were measured through two circular apertures whose size depends on the seeing; the diameters are $2\times$ and $5\times$ the full-width-half-maximum (FWHM) of the point-spread function (PSF). An average correction of the small aperture to the large aperture is computed for bright stars, and this correction is then applied to the smaller aperture magnitudes. 

Stellar catalogues from overlapping images were used to check the photometry. The photometry of two overlapping images was consistent to 0.005 magnitudes for the two $r$-bands, comparable to the quoted accuracy of the PS-DR1 photometry. However, the photometry of the GRI.MP9605 band was found to be good to only 0.02 magnitudes. This is due to two factors. First, the transformation between the Pan-STARRS $g$, $r$ and $i$ filters and our wide filter is complex, with non-negligible scatter. Second, the wide filter causes the PSF to vary, depending on the spectral energy distribution of the stars: bluer stars will have larger PSFs, since seeing is worse at blue wavelengths, and differential chromatic refraction (\S~\ref{sec:dcr}) will stretch some stars away from the horizon.

\subsection{TNO Detection Efficiency}
\label{sec:analysis}

We first consider the total area of sky that OSSOS surveyed.
The on-sky footprint of the eight blocks described in sections~\ref{sec:13BL} through \ref{sec:15BD} (Fig.~\ref{fig:blocks}) is in total $169.62$ deg$^2$. 
The total searched area for OSSOS is 155.3 deg$^2$, with a geometry which is summarized in Table~\ref{tab:pointings}. 
This slightly smaller region accounts for both the true pixel area of the MegaCam field of view, and for the few incompletely searched chips. 
Active pixels fill 91.64\% of the region inside the outer boundary of the MegaCam mosaic: each of the 36 or 40 chips is 2048 x 4612 pixels, and with the 0.18689 \arcsec/pixel plate scale, the field of view is 0.9164 or 1.018 deg$^{2}$. 
Rare pipeline failures occurred in 1-2\% of cases. They were caused by bright stars or large galaxies that elevated the background light levels and reduced point source detections to a level that precluded our automated PSF construction process. This rate is consistent with the sky coverage of the background contaminants.
The effects are quantified by the ``filling factor'' in Table~\ref{tab:characterization}.

The TNO discovery pipeline and analysis techniques for OSSOS are exhaustively described in \S~4 and 5 of \citet{Bannister:2016a}. 
The depth of each discovery triplet of images was quantified by planting PSF-matched sources into copies of each image, at magnitudes $m_r = 21.0 - 25.5$. 
We planted 44k sources in each of 13AE and 13AO \citep{Bannister:2016a}, of order 2 million sources into 13BL and into 14BH, 700k sources into 15AP, and 400k in 15AM and all 15B blocks. 
We found empirically that about 10k sources per CCD were required to a) provide the resolution needed to precisely determine the efficiency as a function of motion rate and b) accurately measure the pipeline's ability to correctly determine the true flux of a source as a function of magnitude. 
(Sources were planted and candidates recovered in iterative loops to avoid planting saturation).
This substantial increase in the number of artificial sources meant that we could not vet 100\% of all artificial candidates \citep[as was done in][]{Bannister:2016a}. 
Instead we audited a subset of about 1000 artificial candidates per field; about 20,000 per block.  
For every frame examined, we performed an audit of the planted sources that we used in calibrating the detection system.  
These audited sources were treated identically to the detections from the actual data frames and were used to verify our planting and search processes.  
To achieve a precise measurement of the structure of the detection threshold and the fill factor, we use the full sample of planted sources.  
We conducted tests to ensure that the vetting process did not introduce an additional bias by examining all planted sources for some blocks.  
These tests revealed that above about the 30\% detection threshold \citep{Petit:2004}, the vetting process does not remove valid detections. 
Our quantification of our false-positive rate is detailed in \citet[\S~5.1]{Bannister:2016a} and was 0 for 13AE and 13AO; in the later 6 blocks we also have a negligible false positive rate. 
Our false-negative rate is $\sim 0.75$\%.
However, the process of vetting all artificial sources did introduce fatigue, which can affect the vetting.
The increase in the number of planted sources and subsequent alteration of our review process was the only material change in our detection and tracking process.

OSSOS provides precisely quantified TNO detection efficiencies. The rate of recovery of sources planted into the images as a function of magnitude is modelled by the functional form given in \citet{Bannister:2016a}:
\begin{equation}
\label{eq:etasquare}
\eta(m) = \frac{\eta_o - c (m - 21)^2}{1 + \exp{\left(\frac{m-m_0}{\sigma}\right)}}
\end{equation}
where $\eta_o$ is the peak efficiency, $c$ is the
strength of the quadratic drop, $\sigma$ the width of the transition from a quadratic to an exponential form and $m_0$ the approximate magnitude of transition.
The peak efficiency $\eta_o$ is roughly the efficiency at $m = 21$, in the case where $\exp((21-m_{0})/\sigma) << 1$.
A few of the efficiency functions could not be fit correctly with the functional form of Eq.~\ref{eq:etasquare}, as the simplex minimisation did not converge with the square function. They were instead fit with a double hyperbolic tangent: 
\begin{equation}
\label{eq:Jones06}
\eta(m) = \frac{\eta_o}{4} \left[1 - \tanh\left(\frac{m - m_0}{\sigma}\right)\right] \left[1 - \tanh\left(\frac{m - m_0}{c}\right)\right]
\end{equation} 
as described in \citet{Jones:2006jl}, and are marked as such in Table~\ref{tab:characterization}.
The efficiency function is determined using the intrinsic flux from the source, i.e. the planted source magnitude.  
The variation in the measured flux from the intrinsic value (caused by noise in the measurement process) is modelled as a function of source flux, for each block. This variation is accounted for during our survey simulation process \citep[see][Appendix A]{Bannister:2016a}. 
For completeness, we list $m_{50\%}$, the $r$-band magnitude at which the efficiency function drops to 50\%, which was often used in earlier surveys. 
However, the key limiting magnitude used throughout OSSOS is instead $m_{limit}$, the slightly deeper, but no fainter than $m_{40\%}$ \citep{Petit:2004} {\it characterization limit}: per \citet{Bannister:2016a}, "the magnitude above which we have both high confidence in our evaluation of the detection efficiency, and find and track all brighter objects". 
The efficiency function parameters are given for all of OSSOS in Table~\ref{tab:characterization}, and are distributed in machine-readable form within our Survey Simulator (see \S~\ref{sec:framework}).

\startlongtable
\begin{deluxetable}{ccccccc}
\tablecolumns{7}
\tabletypesize{\scriptsize}
\tablecaption{Characterization limits for each survey sky region (block) of the Outer Solar System Origins Survey (OSSOS) \label{tab:characterization}}
\tablehead{\colhead{Motion rate ($''\!$/hr)} & \colhead{$\eta_o$} & \colhead{$c$} & \colhead{$\sigma$} & \colhead{$m_0$} & \colhead{$m_{50\%}$} & \colhead{$m_{limit}$} }
\startdata
\cutinhead{13AE}
0.5--8.0   &  0.8877 &  0.0276 &  0.1537 & 24.1423 & 23.95 & 24.09 \\
8.0--11.0  &  0.8956 &  0.0231 &  0.1571 & 24.0048 & 23.86 & 23.85 \\
11.0--15.0 &  0.8658 &  0.0212 &  0.1555 & 23.8811 & 23.74 & 23.73 \\
\cutinhead{13AO}
0.5--7.0   &  0.8414 &  0.0205 &  0.1107 & 24.5497 & 24.36 & 24.40 \\
7.0--10.0  &  0.8776 &  0.0188 &  0.1217 & 24.4187 & 24.28 & 24.26 \\
10.0--15.0 &  0.8639 &  0.0188 &  0.1453 & 24.2575 & 24.10 & 24.10 \\
\cutinhead{13BL}
0.5--2.5   &  0.8524 &  0.0168 &   0.1489 & 24.5213 & 24.35 & 24.45 \\
2.5--8.0   &  0.8883 &  0.0133 &   0.1453 & 24.4793 & 24.36 & 24.42 \\
8.0--12.0  &  0.8841 &  0.0092 &   0.1610 & 24.3311 & 24.24 & 24.22 \\
12.0--15.0 &  0.8683 &  0.0106 &   0.1544 & 24.2243 & 24.12 & 24.10 \\
\cutinhead{14BH}
0.5--2.0   &  0.9359 &  0.0088 &  0.1769 &  24.6522 & 24.57 & 24.66 \\
2.0--6.0$^\star$ &  0.9025 &     0.2820  & 0.7796 & 24.7967 & 24.54 & 24.67 \\
6.0--8.0   &  0.8997 &  0.0101 &  0.1713 & 24.5681 & 24.45 & 24.55 \\
8.0--10.0  &  0.8932 &  0.0113 &  0.1663 & 24.5113 & 24.40 & 24.39 \\
10.0--12.0 &  0.8784 &  0.0113 &  0.1670 & 24.4454 & 24.32 & 24.32 \\
12.0--15.0 &  0.8663 &  0.0140 &  0.1674 & 24.3540 & 24.21 & 24.21 \\
\cutinhead{15AP}
0.5--2.0   &  0.9479 &  0.0117 &  0.1837 & 24.7797 & 24.66 & 24.77 \\
2.0--5.0   &  0.9270 &  0.0131 &  0.1780 & 24.7708 & 24.64 & 24.80 \\
5.0--8.0   &  0.8990 &  0.0124 &  0.1744 & 24.6999 & 24.56 & 24.66 \\
8.0--10.0  &  0.8775 &  0.0125 &  0.1749 & 24.6332 & 24.49 & 24.49 \\
10.0--12.0 &  0.8600 &  0.0124 &  0.1771 & 24.5627 & 24.41 & 24.41 \\
12.0--15.0$^\star$ &  0.8258 &    0.3112 & 0.9109 & 24.6297 & 24.28 & 24.32 \\
\cutinhead{15AM}
0.5--2.0   &  0.9382 &  0.0107 &  0.1561 & 24.9223 & 24.82 & 24.91 \\
2.0--6.0   &  0.9168 &  0.0127 &  0.1490 & 24.9023 & 24.78 & 24.87 \\
6.0--8.0   &  0.8903 &  0.0131 &  0.1485 & 24.8419 & 24.71 & 24.79 \\
8.0--10.0  &  0.8810 &  0.0136 &  0.1462 & 24.7839 & 24.64 & 24.63 \\
10.0--12.0 &  0.8654 &  0.0145 &  0.1482 & 24.7113 & 24.56 & 24.55 \\
12.0--15.0 &  0.8455 &  0.0149 &  0.1518 & 24.6220 & 24.45 & 24.45 \\
\cutinhead{15BS}
0.5--2.0   &  0.9593 &  0.0066 &  0.2064 &  25.3835 &25.29 & 25.39 \\
2.0--6.0   &  0.9353 &  0.0072 &  0.1985 &  25.3351 &25.23 & 25.15 \\
6.0--8.0   &  0.9165 &  0.0075 &  0.1864 &  25.2459 &25.13 & 25.23 \\
8.0--10.0  &  0.9079 &  0.0092 &  0.1820 &  25.1959 &25.06 & 25.07 \\
10.0--12.0 &  0.8965 &  0.0091 &  0.1706 &  25.1040 &24.98 & 24.97 \\
12.0--15.0 &  0.8711 &  0.0104 &  0.1697 &  24.9981 &24.84 & 24.85 \\
\cutinhead{15BT}
0.5--2.0   &  0.7794 &  0.0096 &   0.1543 &25.0931 & 24.88 & 25.00 \\
2.0--6.0   &  0.7337 &  0.0070 &   0.1764 &25.0357 & 24.79 & 24.97 \\
6.0--8.0   &  0.7045 &  0.0072 &   0.1792 &24.9767 & 24.69 & 24.85 \\
8.0--10.0  &  0.6822 &  0.0068 &   0.1694 &24.9149 & 24.62 & 24.68 \\
10.0--12.0 &  0.6802 &  0.0099 &   0.1690 &24.8709 & 24.50 & 24.59 \\
12.0--15.0 &  0.6548 &  0.0096 &   0.1756 &24.7751 & 24.35 & 24.47 \\
\cutinhead{15BC}
0.5--2.0   &  0.9333 &  0.0122 &   0.1300 & 24.8779 &24.78 & 24.86 \\
2.0--6.0$^\star$    &  0.8783 &   0.2351 &  0.7915 & 24.9270 & 24.68 & 24.78 \\
6.0--8.0   &  0.8632 &  0.0108 &  0.1364 &  24.7044 & 24.58 & 24.67 \\
8.0--10.0  &  0.8440 &  0.0125 &  0.1456 &  24.6313 & 24.49 & 24.48 \\
10.0--12.0 &  0.8126 &  0.0114 &  0.1313 &  24.5264 & 24.38 & 24.37 \\
12.0--15.0 &  0.7847 &  0.0129 &  0.1457 &  24.3944 & 24.21 & 24.22 \\
\cutinhead{15BD}
0.5--2.0   &  0.9526 &  0.0072 &  0.1517 & 25.2270 & 25.15 & 25.23 \\
2.0--6.0   &  0.9305 &  0.0093 &  0.1713 & 25.1766 & 25.06 & 25.15 \\
6.0--8.0   &  0.9093 &  0.0111 &  0.1559 & 25.0603 & 24.94 & 25.03 \\
8.0--10.0  &  0.8898 &  0.0116 &  0.1607 & 24.9649 & 24.82 & 24.82 \\
10.0--12.0 &  0.8778 &  0.0118 &  0.1532 & 24.8430 & 24.71 & 24.70 \\
12.0--15.0 &  0.8522 &  0.0144 &  0.1584 & 24.6762 & 24.50 & 24.51 \\
\enddata
\tablecomments{See text for details of $\eta_o, c, \sigma, m_0, m_{50}, m_{limit}$. Fits were made with Eq.~\ref{eq:etasquare}, except those marked with $^\star$, which were fit with Eq.~\ref{eq:Jones06}.
}
\end{deluxetable} 

Detectability and the effective area of survey sky coverage are a function of the distant minor planets' sky motion rate.
For each block, we report multiple efficiency curves in Table~\ref{tab:characterization}, each for a range of rate of motion.  
These different efficiency curves account for the changes in effective area, as faster moving objects are more likely to shear off a given chip between the exposures of the discovery triplet. 
This produces a variation in area coverage as a function of rate, which is only $\sim 2\%$ less for motion rates $>10\arcsec$/hr.
The flux limit for each rate is also different: trailing of fast moving objects makes the flux sensitivity poorer, as the source is spread over more pixels in each exposure, pushing faint sources into the noise. 
We set an upper limit of $15\arcsec$/hr as the survey's focus is on discoveries at heliocentric distances $\gtrsim 10$ au; at rates faster than this, the object would also feature significant trailing within an exposure, causing reduced detection efficiency (partly due to reduced signal to noise and partly due to our pipeline being optimised for roughly circular sources). 
The survey simulator (\S~\ref{sec:framework}) accounts for this variation in sensitivity as a function of TNO sky motion rate.  

\subsubsection{TNO tracking efficiency}
\label{sec:tracking}

Our detected and tracked sample is a fair representation of the orbital distribution in the Kuiper belt.  
Of the 840 TNOs brighter than the characterization limit, 99.8\% were recovered outside of their discovery triplet and tracked to fully classifiable arcs (\S~\ref{sec:orbitquality}). 

All of the $\sim 37,000$ tracking observations that comprise the arcs were individually inspected. Aperture photometry was measured with \texttt{daophot} \citep{Stetson:1987} with a typical 4-5 pixel aperture, corrected to a 20-25 pixel aperture using bright-star aperture corrections, and the sky measured in a 10 pixel wide annulus with an inner edge of about 30 pixels. While all images are calibrated to Panstarrs/Gaia (\S~\ref{sec:photcalib}) and standard MPC flags\footnote{\url{minorplanetcenter.net/iau/info/ObsNote.html}} were assigned throughout to indicate any photometric irregularities, the arc measurements were optimized for astrometry rather than precision photometry. Thus, care should be taken if using the bulk arcs to infer TNO light curves or variability.

The superlative tracking fraction removes ephemeris bias from the survey.
Only two detections could not be tracked, both with estimated distances interior to 20 au: \texttt{o5p002nt} and \texttt{o5s03nt} in Table~\ref{tab:discoveries}~(characterized).
These two detections are brighter than the characterization limit and have rates of motion within the survey limits, i.e. they are characterized discoveries.
The loss of these two objects is not magnitude dependent. 
Both moved onto the corner of their blocks only during the night of the discovery triplet, so there were no additional observations with which to track them.  
The loss of these two objects is accounted for using the rate-dependent tracking fraction parameter in our survey simulator (\S~\ref{sec:framework}), which applies an accurate reproduction of the OSSOS survey tracking success.

\subsubsection{Sensitivity as a function of heliocentric distance}

The maximum distance sensitivity of a survey is set by the temporal spacing of the discovery exposures, the elongation of the target field and the length of the exposures. 
For OSSOS, it varies across the range of blocks: each has a unique set of delays between our three exposures and a unique field elongation. 
Additionally, the trailing of sources results in poorer sensitivity to fast-moving sources (i.e. inside 20 au distance).
To determine our distance sensitivity, we examine the efficiency curves for the slowest rate bin for each block (Table~\ref{tab:characterization}).
The 0.5\arcsec/hr threshold to which we search implies a detectable distance limit of $\sim300$ au. 
However, taking flux into account, our 50\% threshold for large $H_r = 4$ TNOs ranges between a detectable distance of 100 au for 13AE and 130 au for 15BD.
Having considered temporal spacing/elongation and flux, we additionally take into account the steep size distribution of TNOs. 
In our sample we have few detections beyond 60 au. 
Our most distant detection was at 82.5 au, $H_r = 5.6$, in 15BC (Fig.~\ref{fig:pointings}). 
One might be inclined to infer that this indicates an intrinsic lack of TNOs between 60 au (our decrease in detections) and 100 au (our practical limit for detection of TNOs with $H_r > 4$). 
However, in the entire OSSOS survey we have only 2\% of our sample drawn from objects with $H_r < 6$, and only one object with $H_r < 4$ (Table~\ref{tab:discoveries}). 
Thus, our lack of distant $>60$ au detections is more a manifestation of the steep luminosity function of TNOs and the area coverage of OSSOS, rather than of our flux or rate of motion sensitivity.

\subsubsection{Sensitivity to retrograde orbits and interstellar objects.}

The OSSOS detection pipeline efficiency is independent of the nature, sun-bound or interstellar, of the trajectories of the minor planets in the field of view.  
The detection process is driven only by the angular rate and direction of each source on the sky. 
We consider in turn the cases of retrograde orbits and interstellar trajectories.

For both retrograde and prograde orbits bound to the Solar System, the component of reflex motion from the Earth's orbital velocity dominates the direction of sky motion in our imaging.
Orbits that are retrograde will move about 10\% faster on the sky than prograde TNOs at the same distance. 
Our coverage both before and after the discovery night provides tracking that covers all possible orbit motion ranges, removing nearly 100\% of ephemeris bias.
99.8\% of our detected sources brighter than our characterization limit were tracked to full orbits, and the 0.2\% tracking loss rate of very close objects was not due to orbit assumptions (\S~\ref{sec:analysis}).
Thus, although none of the minor planets that we detected were on retrograde orbits, we can conclude that if such objects had been present in our fields and fell inside our rate and direction cuts, we would have detected {\it and} tracked them.
We note that our algorithm continues that of CFEPS, which detected the first retrograde TNO \citep{Gladman:2009kv}.
Because we chose the ensemble of rates and directions to cover {\it all} bound Solar System orbits beyond $\sim10$~au, we are confident that no retrograde orbits brighter than our survey limits were within our detection volume.
An upper limit for the retrograde population could be created, which would be highly dependent on the assumed distribution of $a$, $q$, and $H$, and we defer that to future work.

OSSOS has less sensitivity to unbound interstellar interlopers like 1I/`Oumuamua \citep{Meech:2017}.
Interstellar objects have very large heliocentric velocity vectors, cf. the 26 km/s of 1I/`Oumuamua. 
For most of the part of their inbound and outbound trajectory when they are close enough to be brighter than our flux limit, unbound orbits could be moving in almost any direction on the sky. 
Thus, interstellar objects would mainly be outside our rate cuts, though could be detectable if within our rate cuts.

\section{Discoveries}
\label{sec:discoveries}

OSSOS detected 949 objects within the eight survey blocks.
We present the catalogue of their discovery and orbital properties in three machine-readable tables, which are separated by their quality of usefulness as samples for modelling the structure of the outer Solar System. The content of each table is summarized in Table~\ref{tab:discoveries}.
In the following section, we discuss the quality of our TNO orbits and their classification into different dynamical populations, including the small fraction of our discoveries that were linked to previously known objects. 

\begin{deluxetable}{lcl}
\tablecolumns{3}
\tablecaption{Description of the survey discovery catalogues \label{tab:discoveries}}
\tablehead{\colhead{Column Name} & \colhead{Unit} & \colhead{Description} }
\startdata 
\sidehead{Table: Characterized OSSOS minor planets (840 rows)}
\sidehead{Table: Uncharacterized OSSOS detections (109 rows)}
\sidehead{Table: Ensemble of characterized minor planets from four surveys (1142 rows)}
\hline
cl &	& Orbital population/class \\
p  &	& Additional class-dependent detail \\
j &	& Resonant object is in the $j:k$ resonance \\
k &	& Resonant object is in the $j:k$ resonance \\
sh  &	& Orbit classification status (secure/insecure) \\   
object  &	& Survey object designation \\
mag & mag & mean magnitude during the discovery triplet,\\&&excluding flagged observations \\
mag$_e$ & mag & Uncertainty in mag \\
Filt &	& Filter used in discovery observation \\
$H_{sur}$ & mag	& Surmised absolute magnitude H, in discovery filter\\
dist  &	au & Object distance at discovery \\
dist$_e$    & au	& Uncertainty in dist \\
N$_{obs}$   &	& Number of observations available \\
time  & years & Length of measured orbital arc \\
av$_{xres}$ & \arcsec	& Mean orbit-fit residual, R.A. \\
av$_{yres}$ &	\arcsec & Mean orbit-fit residual, Decl. \\
max$_{x}$  & \arcsec	& Maximum orbit-fit residual, R.A. \\
max$_{y}$  & \arcsec & Maximum orbit-fit residual, Decl. \\
$a$          &	au & Semimajor axis \\
$a_e$        &	au & Uncertainty in $a$ \\
$e$        &	& Eccentricity \\
$e_e$      &	& Uncertainty in $e$ \\
$i$      &	\degr & Inclination to the ecliptic \\
$i_e$      & \degr 	& Uncertainty in $i$ \\
$\Omega$ &	\degr  & Longitude of ascending node \\
$\Omega_e$    &	\degr & Uncertainty in $\Omega$ \\
$\omega$   & \degr 	& Argument of perihelion \\
$\omega_e$      & \degr & Uncertainty in $\omega$ \\
$T_{peri}$    & days	& Modified Julian Date of osculating perihelion passage \\
$Tperi_e$   & days	& Uncertainty in tperi \\
R.A.   &	\degr & Right ascension (J2000) at mean time of discovery \\
Decl.     &	\degr & Declination (J2000) at mean time of discovery\\
JD        &	& Central Julian Date of first discovery image\\
rate & \arcsec/hr & Angular rate of sky motion at discovery \\
MPC	 & & Minor Planet Center object designation (packed format) \\
\enddata 
\tablecomments{
All orbital elements are J2000 ecliptic barycentric coordinates. MPC packed format is defined at \url{http://www.minorplanetcenter.net/iau/info/PackedDes.html}. The three machine-readable discovery catalogue tables are in the Supplementary Materials.
}
\end{deluxetable}

\subsection{Discoveries to use for testing models}

The OSSOS primary discovery set is 840 TNOs, given in the catalogue described in Table~\ref{tab:discoveries}: {\it characterized} OSSOS minor planets. These minor planets were found at brightnesses and rates of sky motion that mean they have {\it fluxes brighter than our tracking limit}, with well-quantified detection efficiencies. (This includes two close objects that were not tracked: see \S~\ref{sec:analysis}). The characterized discoveries are the set that should be used for modelling Solar System structure. We summarize their orbital populations in Table~\ref{tab:totals}.

\begin{deluxetable}{rrrrl}[h]
\tablecolumns{5}
\tablecaption{Orbital populations of the characterized discoveries of the Outer Solar System Origins Survey \label{tab:totals}}
\tablehead{\colhead{Dynamical class} & \colhead{$a$ (au)} & \colhead{Secure} & \colhead{Insecure} & \colhead{Comments} }
\startdata 
Jupiter-coupled    & &   5  &  0 & $q<7.35$ au and $T_j<3.05$	\\
Centaur			   & $a<30$ &  15  &  0 & Insecure one is \texttt{nt}	\\  
scattering     	   & &  29  & 9	& $\delta a > 1.5$ au in 10 Myr	\\
inner-belt         & $a<39.4$ &  11  &  0	&	\\
main-belt          & $39.4 < a < 47.7$ & 391  & 30	&	\\
outer-belt         & $a > 47.7$&   4  &  0	& $e<0.24$	\\
detached component & $a>47.7$ &  10  & 21	&$e>0.24$	\\ 
Total non-resonant & & 465  & 60 &  +2 \textit{nt} discoveries \\
\cutinhead{Neptune mean-motion resonant ($n$:$m$) ordered by increasing $a$}
1:1  & 30.1 & 4	& 0	& Neptune trojan \\
4:3  & 36.4 &	10  & 0 &	\\
3:2  & 39.4 &	131 & 1	&	\\
5:3  & 42.3 &	13 & 1 &	\\
7:4  & 43.7 &   31	& 8	&	\\
9:5  & 44.5 & 1 & 1 & \\
2:1  & 47.7 &	34 & 0	&	\\
5:2  & 55.4 &	17 & 3 & \citet{Volk:2016} \\
3:1  & 62.5 & 6	& 1	&	\\
7:2  & 69.3 & 1	& 1	&	\\
4:1  & 75.7 &	1 & 2	&	\\
9:2  & 81.9 &	0 & 1 & \citet{Bannister:2016AJ_OSSOSIV} \\
5:1  & 87.9 & 0 & 1 & \\ 
9:1  & 130.1 &	2 &	0 & \citet{Volk:2017inprep}	\\
Others & & 19 & 23 &	\\
Total resonant &       & 270 & 43  &	\\
\hline
Total (all)     &     & 735 & 103 &	+2 \textit{nt} discoveries \\
\enddata 
\tablecomments{Classifications similar to that in \citet{Gladman2008}, with the difference that resonant status is declared if the best-fit clone resonates even if the two extremal clones do not. We check all objects thoroughly for mean-motion resonances with Neptune; we additionally check low-$a$ objects for 1:1 mean-motion resonances with Saturn and Uranus. The semi-major axes given for resonances are approximate.}
\end{deluxetable} 

\subsection{Discoveries with poorly determined discovery efficiency and some incomplete tracking}

For completeness, we list all 109 remaining detections in the catalogue referenced in Table~\ref{tab:discoveries} as {\it uncharacterized} OSSOS detections. 
In this machine-readable catalogue, there are three TNOs that have detection efficiencies that are quantifiable, but differ from the efficiencies provided here for the main Survey. 
They are designated with the prefix \texttt{Col3N} and were fast-moving discoveries in the incomplete first attempt at a discovery triplet on the 14BH block in 2013 (\S~\ref{sec:14BH}). 

Table~\ref{tab:discoveries} ({\it uncharacterized}) also catalogues 106 TNOs from all blocks that are designated with a prefix {\tt u}. 
These exceptionally faint objects have a detection efficiency that is poorly quantified, as their faintness at discovery placed them below their block's rate-of-motion-dependent {\it characterization limit}. 
The uncharacterized objects are severely afflicted by unquantifiable observational biases.
The lack of quantified detection efficiency means that the only aspect of the tracked, uncharacterized TNOs that we can be sure of is their orbits. 
Of the uncharacterized TNOs, 24 could not be tracked beyond the $\sim 2$-hour arc of their discovery triplet; these are designated with a suffix {\tt nt}.
It is possible that some of the untrackable {\tt nt} sample may actually be false positives; see Sec. 5.1 of \citealt{Bannister:2016a} for an evaluation of the false positive rate of our analysis.
We recommend exceptional caution in any use of the Table~\ref{tab:discoveries} uncharacterized TNOs, and strongly advise against simply combining them with the characterized discoveries when modelling Solar System structure.

\subsection{Orbit quality and population classifications}
\label{sec:orbitquality}

To fully exploit the discovery of a TNO we must know both its discovery circumstances and its precise orbit.
Extensive tracking is often required before such precise orbits can be determined; this is especially critical for resonant objects.  
The resonant orbits require extreme orbital precision (fractional semi-major axis uncertainty roughly $<$0.01\%) in order to determine the resonance libration amplitude, which is a diagnostic of the resonance capture mechanism. 
An additional classification to those in \citet{Bannister:2016a} is for Jupiter-coupled objects; these are defined by a combination of perihelion distance and Tisserand parameter:
\begin{equation}
T_{J} = a_{J}/a_{TNO} + 2\sqrt{ (a_{J}/a_{TNO})(1-e_{TNO}^2)  }\cos(i_{TNO})
\end{equation}
We do not attempt orbital classifications for the {\tt nt} discoveries or for objects with an arc only within their discovery lunation.

As described in \citet{Bannister:2016a}, we classify the TNOs into different dynamical populations \citep{Gladman2008}, based on a 10 Myr integration of clones of the best-fit orbit and two extremal orbits, which are determined from the available astrometry using the orbit fitting method of \citet{Bernstein:2000p444}.
When all three clones show consistent dynamical behaviour, we declare the object's classification to be {\it secure}.
Precisely determining how long an observing arc, or the frequency of observations, that are required to secure an orbit is not strictly possible because it depends on the TNO's dynamical behaviour.  
The trans-Neptunian region is riddled with chaos and complex boundaries in orbital phase space.  
Orbits near the boundaries of resonances must be very precisely determined before the dynamical evolution can be accurately presented.  
In contrast, secure classifications for orbits farther from the resonance boundaries can be achieved with less precision.
We do not know in advance where a given object's orbit is relative to these boundaries. 

The high-precision and well-sampled tracking of OSSOS has allowed 734 of the 840 characterized orbits to be securely classified with the arcs we present here. 
This is summarized by population in Table~\ref{tab:totals}.
Additional arc was required to further refine the orbits for some insecurely-classified objects. 
The orbits of the \texttt{o3e} and \texttt{o3o} TNOs that were initially reported in \citet{Bannister:2016a} now have an additional two oppositions of more sparsely sampled OSSOS arc, and accordingly have updated classifications. 
Once an object was securely classified we discontinued target tracking (occasionally, further serendipitous observations may have occurred, which we measure and include; see \S~\ref{sec:observations}). 

At the completion of the 2013--2017 observing, 103 characterized discoveries with well-observed arcs remain insecurely classified.
The chaotic nature of the resonance boundaries means these objects' dynamical classifications might never become secure, because they transition between resonant and non-resonant behaviour during the 10 Myr of the classification integration. 

\subsection{Previously discovered TNOs}
\label{sec:PD}

Fewer than 3\% of the OSSOS discoveries were previously known, largely because of the greater depth of OSSOS compared to previous wide-field surveys.
We designate 23 TNOs in Table~\ref{tab:discoveries} (characterized OSSOS) with the suffix \texttt{PD}, which indicates that during their tracking (\S~\ref{sec:observations}) we made use of available pre-OSSOS astrometry to assist in securing their orbits. 
Pre-OSSOS astrometry was located in two ways.
After the discoveries from each block were made, we checked against the predicted location of all Minor Planet Center-listed TNOs that were within the sky coverage. 
We also tested for linkages to the short-arc discoveries by CFEPS \citep{Petit:2011p3938}. 

There are OSSOS objects with pre-OSSOS data that we do not designate as \texttt{PD}.
Several OSSOS discoveries were subsequently given MPC designations earlier than 2013-2015 on submission, due to linkages that the MPC was able to provide to unlisted or exceptionally sparse earlier astrometry: e.g. \texttt{o3l28} (2006 QE$_{181}$). 
These do not have the suffix \texttt{PD}.
Additionally, some objects were independently discovered during our survey. 
\texttt{o3l39} (2016 BP$_{81}$) was found in {\it Kepler} space telescope observations of the K2 Campaign 8 superstamp field of Uranus \citep{Molnar:2017arXiv}. 
\texttt{o3l18} (2010 RE$_{188}$) was reported by Pan-STARRS \citep{Holman:2015,Weryk:2016}.
The orbits of these TNOs were already securely classified with our observed arcs by the time the other astrometry came available, and we do not designate them with \texttt{PD}.

The high cadence and high astrometric precision of OSSOS maintains our sample uniformity across all our discoveries, including those designated \texttt{PD}.
We tested for the effect of the prior astrometry on our classifications by integrating our arcs both with and without the earlier observations.
Our orbital classifications are independent of the possibility that an object might be a \texttt{PD}: no classifications changed in class, though a few shifted between secure and insecure.
Note that none of the \texttt{PD} objects are on high-perihelion, exceptionally large-$a$ orbits (\S~\ref{sec:extreme}), where instead the observed length of arc would become the dominant source of orbital uncertainty.
The \texttt{PD} TNOs are consistent in quality with the rest of the dataset, rather than exhibiting greater orbital precision.
This held true even in cases where the earlier astrometry extended the orbital arcs to a decade or more in length.

Our linkages to earlier discoveries show that for sparsely sampled arcs, even multi-opposition orbits in the MPC are sometimes misleading.
\texttt{o5p064PD} (2001 FL$_{193}$) is an interesting case.  
The 2001 discovery by the Deep Ecliptic Survey \citep{Elliot:2005ju} was with a pair of observations on March 27 and one on March 28. 
Fourteen years later, 2001 FL$_{193}$ was recovered in three single observations in 2015 by the DECam NEO Survey \citep{Allen:2014DPS}.
Having only six measured points defined 2001 FL$_{193}$'s orbit with high uncertainty, despite its 14-year arc. 
The orbit was a rare low-inclination $a \sim 50$~au outer belt TNO beyond the 2:1 (see Table 6, \citet{Bannister:2016a}).
The OSSOS discovery \texttt{o5p064PD} links back to this previous data. 
However, we find instead that \texttt{o5p064PD} = 2001 FL$_{193}$ is correctly on an $a = 43.1$~au, $e = 0.04$, $i = 1.8\degr$ orbit, which is that of a standard cold classical TNO.

A few MPC-listed objects were nominally within our coverage, but fell in chip gaps during our discovery observations (e.g. 2000 FA$_{8}$ was in a chip gap during the 2015 April 12 discovery observations for the 15AP block). We recovered and reported astrometry where these TNOs were serendipitously included in our tracking observations.

\section{The survey ensemble: a framework for testing models of the Solar System's dynamical history}
\label{sec:framework}

We provide a survey simulator\footnote{\doi{10.5281/zenodo.591435}; ongoing development and the most recent survey files are at \url{https://github.com/OSSOS/SurveySimulator} }, described in detail in \citet[][\S~5.2]{Bannister:2016a}, which can assess model trans-Neptunian populations against the OSSOS orbit sample. 
With this piece of software one can take a model of the outer solar system and submit it to an observed survey's detection characterization. 
The model needs to provide a set of small outer solar system bodies defined by their orbital elements and physical characteristics, e.g. intrinsic absolute magnitude and colours. 

The survey simulator allows the user to model multiple well-characterized surveys performed using up to ten different filters, and to model populations of objects that may vary in color \citep{Bannister:2016a}. 
Each model object given to the survey simulator needs to be specified with its intrinsic absolute magnitude, given in a band-pass filter, $X$, and an array of colors between the various survey filters that each survey used and the reference $X$-band. 
Thus the variation of colors inside and across the various dynamical populations can be modelled when combining surveys performed through different filters. 

For each object, the survey simulator computes the position in the sky of the object, its apparent motion, and its apparent magnitude, accounting for the Poisson noise in the measured flux. 
The survey simulator models the observing conditions (image quality, attenuation, star crowding) through the use of the detection efficiency function (\S\ref{sec:analysis}). 
Then it decides if the object was detected and tracked, determining what a given survey would have detected, had the outer solar system been the model.
The orbital and physical characteristics of the model ``detections'' can then be compared to the discovered TNO sample to assess the viability of the model. 

In previous version of the survey simulator, we used rectangles with the edges along the RA and DEC axes to describe the footprint of the blocks on the sky, which was appropriate for the RA-Dec grid-aligned survey blocks of CFEPS and A16. 
We had to balance between the number of rectangles (more rectangles slow down the computation) and the accuracy of the simulation (fewer rectangles imply less accurate representation of a non-rectangular shape). 
We now describe the footprint of the blocks by polygons; the areas are given in Table~\ref{tab:pointings}. 
This allows us to accurately represent the shape of the observed region, while still being numerically efficient. 
The shape accuracy is particularly important for large, staggered blocks like those in OSSOS that straddle the plane of the Kuiper belt.

The simulator that we provide here includes four characterized surveys with well-measured discovery biases (Table~\ref{tab:ensemble}): CFEPS, HiLat, A16 and OSSOS, all made with CFHT's MegaCam.
Only the colours $g-X$, $r-X$ and $R-X$ need to be specified to the simulator when modelling this full set of surveys, and only $r-X$ when modelling OSSOS, HiLat and A16.
The ``ensemble'' of surveys covers in total 1224 deg$^2$ of sky, at a wide range of ecliptic latitudes and longitudes, to depths of $m_r = 23.5-25.2$.
There are 1142 TNOs in the ensemble discovery sample, from a wide range of TNO dynamical populations. 
We provide them in the same format as for the OSSOS discoveries (Table ~\ref{tab:discoveries}). They are collated in the catalogue referenced in  Table~\ref{tab:discoveries} as the {\it ensemble}.
This is by far the largest sample with quantified biases currently available for investigating the structure of the trans-Neptunian populations. 
We observed the insecurely classified TNOs from the earlier surveys several times during the observations of OSSOS, which has improved some of the orbital classifications from those at the time of their discovery surveys' publication.
In the 6 cases where TNOs from earlier surveys in the ensemble were rediscovered by OSSOS, they are listed twice in Table~\ref{tab:discoveries}: {\it ensemble}, in both cases with the OSSOS object designation.

Future characterized surveys can easily be added to the simulator. 
We provide an example template in the Supplementary Materials for the formatting of the characterization information and TNO discovery sample. 

\begin{deluxetable}{llr}[h]
\tablecolumns{3}
\tablecaption{The ensemble of surveys described by the OSSOS Survey Simulator. \label{tab:ensemble}}
\tablehead{\colhead{Survey} & \colhead{Bandpass} & \colhead{Characterized Detections}}
\startdata 
CFEPS$^\dagger$ & G.MP9601 & 210\\
HiLat$^\star$ & R.MP9601 & 21 \\
A16$^\sharp$ & R.MP9601   & 77 \\
OSSOS & R.MP9601/R.MP9602 & 840 \\ 
\hline
Total (distinct detections) & & 1142 \\
\enddata
\tablecomments{$\dagger$: \citet{Jones:2006jl,2009AJ....137.4917K,Petit:2011p3938}
$\star$: \citet{Petit:2017ju}
$\sharp$: \citet{Alexandersen:2016ki}. There are 6 characterized detections in CFEPS that were re-found in OSSOS. All detections are provided in the machine-readable catalogue of Table~\ref{tab:discoveries}: ensemble.}
\end{deluxetable}

\section{An Overview of the trans-Neptunian Populations From Characterized Discoveries}
\label{sec:populations}

\begin{figure*}
\plotone{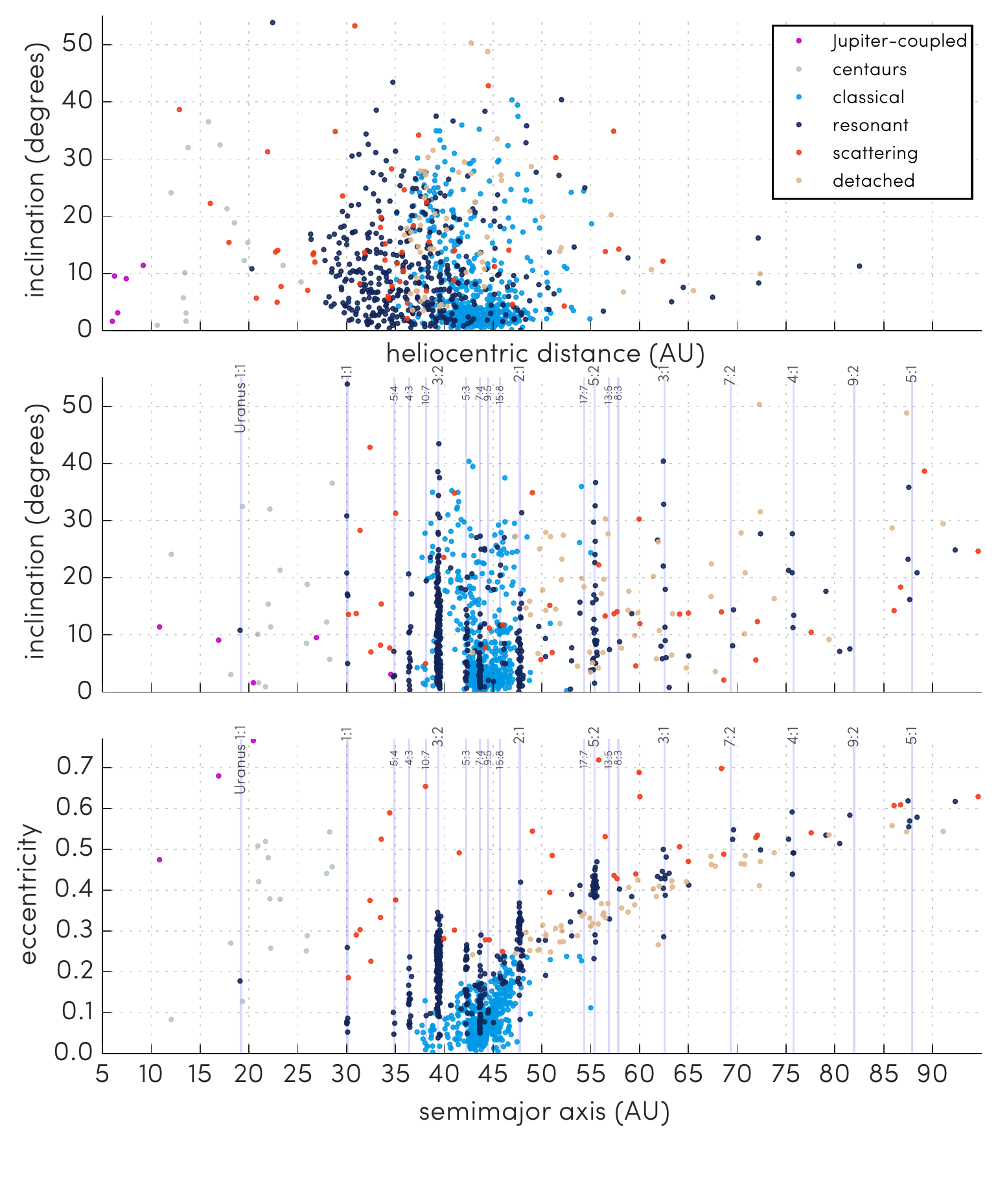}
\caption{Orbital parameters and discovery distances of the 1142 characterized minor planets from the four surveys of the ensemble, OSSOS/CFEPS/HiLat/A16 (Table~\ref{tab:discoveries}: ensemble), displaying the subset with parameters within these $a/e$ and $a/i$ axes ranges. 
All uncertainties are smaller than the point size.
The pale blue vertical lines show the approximate semi-major axis locations of the resonance centres for resonances with detections.
\label{fig:orbdiscoveries}}
\end{figure*}

The orbits of our discoveries reveal new and complex detail in the known trans-Neptunian populations. 
Here we provide a broad overview of the strengths of the OSSOS sample, relative to that present in the MPC. 
Where relevant, we consider the sample available by combining the characterized discoveries of four surveys (all available in the survey simulator): CFEPS, HiLat, A16 and OSSOS, which we refer to as the ``ensemble'' (\S~\ref{sec:framework}).
We focus on the dynamical properties of the characterized ensemble sample (Table~\ref{tab:discoveries}), which is graphically presented in Fig.~\ref{fig:orbdiscoveries} and Fig.~\ref{fig:qi}.
We also draw attention to individual OSSOS objects on unusual orbits (\S~\ref{sec:stirred}, \S~\ref{sec:extreme}). 

\subsection{Minor planets on transient orbits}
\label{sec:interacting}

The Centaur ($a < 30$~au) population and the $q\lesssim 30$ portion of the scattering population undergo strong gravitational interactions with the giant planets, making them an unstable, transient population.
The larger $q$ portion of the scattering population also interact with Neptune and undergo significant orbital evolution as a result, but the more distant encounters allow for longer dynamical lifetimes.
These unstable orbits are a mixture of a slowly-decaying remnant of a vast population emplaced early in the Solar System’s history when the Oort cloud was built, and a steady-state intermediary population leaking from a more stable Kuiper belt or Oort cloud reservoir \citep{Duncan:1997hg,Gomes:2008,Dones:2015}.
OSSOS provides 21 objects classed as Jupiter-coupled and Centaurs, and 38 scattering TNOs. 
\citet{Shankman:2016hi} considered the size distribution of the scattering TNOs using 22 objects: 13 from CFEPS, 2 from A16, and the first 7 OSSOS scattering discoveries. 
The available ensemble sample is now more than doubled. \citet{Shankman:2017thesis,Lawler:2017inprep} find a break in the scattering size distribution, and an intrinsic population of $9 \times 10^4$ for $H_{r} < 8.66$ ($D > 100$~km), with the Centaur population two orders of magnitude smaller at $111^{+59}_{-44}$ \citep{Lawler:chapter}. 

\subsection{The non-resonant TNOs}
\label{sec:belt}

\begin{figure*}
\plotone{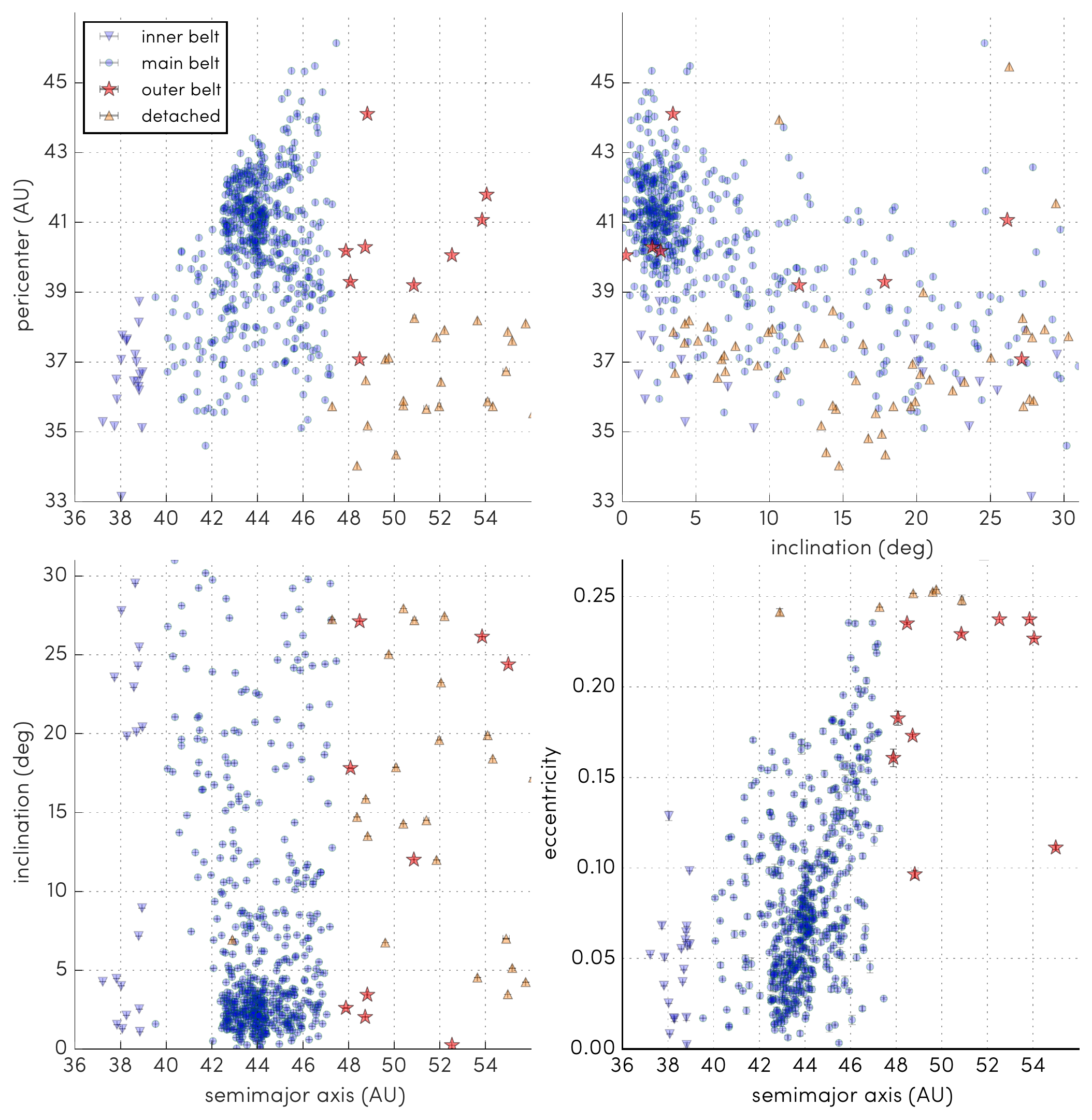}
\caption{The non-resonant Kuiper belt and detached component as observed by the four surveys of the ensemble, OSSOS/CFEPS/HiLat/A16 (Table~\ref{tab:discoveries}: ensemble). 
Inner, main and outer Kuiper belt, and detached TNOs, are distinguished by symbol shape, based on their orbital classifications (\S~\ref{sec:orbitquality}).
The main Kuiper belt (blue circles) shows a strong concentration of low-$i$ orbits with $a\simeq$44~au, a feature known as the {\it kernel} \citep{Petit:2011p3938,Bannister:2016a}.
The outer belt (red stars) continues at low inclinations beyond the 2:1 mean-motion resonance with Neptune ($a \sim 47.7$~au). 
All orbital elements are barycentric, with $1\sigma$ uncertainties shown.
\label{fig:qi}}
\end{figure*}

The OSSOS survey has acute sensitivity to all objects exterior to the orbit of Saturn, and thus provides a strong sampling of the inner-belt component of the TNO population in $37 \lesssim a \lesssim 39$~au. 
OSSOS detected 11 inner-belt TNOs above our characterization thresholds. 
Given our thorough ability to detect objects interior to 37~au, and the complete absence of non-resonant/non-transient objects between $30 < a < 37$~au, we find that the present-day inner belt does not extend inward of $a \sim 37$~au (Fig.~\ref{fig:qi}, top left). 
It is noteworthy that the strong lower perihelia bound of $q \sim 35$~au is consistent across the entire Kuiper belt. 
Only one TNO with a semi-major axis in the region $36 < a < 48$~au has a perihelion dipping just below, to $q=34.6$~au (Fig.~\ref{fig:qi}, top left).
The main structure visible within the inclination distribution in the inner belt is that of the $\nu_8$ secular resonance, which destabilizes orbits in this semi-major axis range that have $i \lesssim 15\degr$ \citep{2009AJ....137.4917K}.
After accounting for this instability zone, the distribution of inner-belt inclinations continues to be consistent with being drawn from the same dynamically excited population as is present in the main belt.  
Further examination of the architecture of the inner belt and its relation to the main belt \citep{Petit:inprep} is likely to provide significant insight into the processes that resulted in the currently observed structure \citep[e.g.][]{Nesvorny:2015et}.

As of 2016 October, the MPC listed 484 TNOs with arcs where orbits as fit showed no resonance, with $42~<~a~<~48$~au and $\sigma_{a} < 5\%$, from surveys other than in the ensemble \citep{Volk:2017}. 
The ensemble presented here has 530 such main-belt objects, of which 421 are from OSSOS, and they have $\sigma_{a} < 0.1\%$: the orbital precision is substantially better, allowing for much more secure dynamical classifications. 
We show these in Fig.~\ref{fig:qi}.
In the main belt, even the first two blocks of OSSOS were able to independently confirm the existence of the cold classical ``kernel'' \citep{Bannister:2016a}, which was first described in \citet{Petit:2011p3938}. 
With the full ensemble sample, the kernel is starkly visible in Fig.~\ref{fig:qi}.
The tight parameter space of the kernel constrains the processes that are viable candidates for its sculpting out of the primordial disk \citep{Gomes:2017}, or its emplacement \citep{Nesvorny:2015ik}.

\subsubsection{The cold classical population extends beyond the 2:1 resonance with Neptune}
\label{sec:stirred}

OSSOS confirmed that the cold classical population extends beyond the 2:1 resonance \citep{Bannister:2016a}, which places a significant new condition on the creation or emplacement of low-inclination orbits in this region.
Only six TNOs are known in this elusive population. 
The ensemble sample has four outer-belt objects that orbit on inclinations of only a few degrees, with $q > 40$~au and semi-major axes beyond the 2:1 mean-motion resonance with Neptune: \texttt{o3e45, o4h47, o5d115PD} from OSSOS, and 2011 US$_{412}$ from A16 (Fig.~\ref{fig:qi}).
The other non-resonant outer-belt TNOs just beyond the 2:1 have more substantial inclinations of 12.7\degr--36\degr (Fig.~\ref{fig:qi}).
We re-discovered two previously known TNOs related to this low-inclination population. 
OSSOS confirms the outer-belt orbit of {\tt o5d115PD}, which is (48639) 1995 TL8, and rules out that of \texttt{o5p064PD} (2001 FL193): it is instead a standard main-belt cold classical (see \S~\ref{sec:PD}).
The others listed in the MPC are 2003 UY$_{291}$ \citep{Gladman2008} and 2012 FH$_{84}$ \citep{Sheppard:2016jy}.
We report a new member of the outer belt.
With $i = 3.4\arcdeg$ and $a = 48.83$ au, $q = 44.10$ au, \texttt{o4h47} is indisputably linked to the cold classical population.
It has the lowest eccentricity and highest perihelion yet seen for a cold classical beyond the 2:1 (see Table~6, \citealt{Bannister:2016a}).

\subsubsection{The detached component}
\label{sec:detached}

The $a > 50$ au OSSOS sample is particularly valuable due to the exceptional certainty of its orbit classifications. 
In the range $50 \leq a \leq 150$ au, there are 135 apparently non-resonant TNOs with $da/a < 5\%$ orbits in the MPC from other surveys. This contrasts with 76 in the ensemble, 48 of which are from OSSOS.
A major non-resonant $a > 50$ au population are on orbits that are {\it detached}: stable, with large $a$ and $e$, but with perihelia large enough that they are not now strongly interacting with Neptune; the exact perihelion distance required to prevent significant orbital evolution due to distant encounters with Neptune depends on semi-major axis, so classification as `detached' can only be determined by numerical integration \citep{Lykawka:2007ff,Gladman2008}.
OSSOS provides 31 TNOs orbiting in the detached component (those with $a$ below the limits of the axes are shown in Fig.~\ref{fig:orbdiscoveries}).
The fractional increase in the non-resonant $a > 50$ au sample from OSSOS is less pronounced than it was for the non-resonant Kuiper belt. 
However, the OSSOS sample is far more secure than the MPC TNOs in this range; it is entirely possible that many objects listed in the MPC are inaccurately classified as non-resonant due to their low-precision orbits.
Additionally, the ensemble sample has well-characterized biases, which is particularly important for the large-$a$ population.

\subsection{Resonant TNOs}
\label{sec:resonant}

OSSOS provides a substantive additional sample of TNOs in mean-motion resonances with Neptune (Fig.~\ref{fig:orbdiscoveries}).
The four new and secure Neptune Trojans from OSSOS exhibit the known wide range of inclinations of this population \citep{Chiang:2003hb, Sheppard:2006et, Sheppard:2010, Parker:2013trojan, Alexandersen:2016ki}, discussed further in \citet{Lin:inprep}. 
A spectacular 132 plutinos (3:2 resonance) dominate the resonant detections of OSSOS; combined with the survey simulator, the detailed de-biased distribution will provide the information needed to study and constrain resonance capture conditions and plutino mobility over the age of the Solar System (Volk et al., in prep.).

Observational biases make the TNOs in more distant resonances progressively harder to detect, as they spend a larger fraction of each orbit at greater heliocentric distances, where they are fainter.
For example, OSSOS found only 34 TNOs in the 2:1 resonance. 
However, early OSSOS findings support earlier findings that the total population in the more distant resonances may rival or exceed those of the non-resonant Kuiper belt \citep{Gladman:2012ed}.
\citet{Volk:2016} found that the OSSOS \texttt{o3e} and \texttt{o3o} discoveries imply that the 5:2 resonance ($a \sim 55$~au) is much more heavily populated than had previously been suggested from migration models.
The significant number of objects OSSOS has observed in some of the resonances will enable detailed modelling of their dynamical structure and relative populations, providing detailed constraints on Neptune's migration history.
For example, the \texttt{o3e} and \texttt{o3o} discoveries confirmed that the objects in the 2:1 have a colder inclination distribution \citep{Gladman:2012ed} than the population of objects in the 3:2 \citep{Volk:2016}.
This is now visually obvious in Figure~\ref{fig:orbdiscoveries}, and is discussed further in \citet{Chen:inprep}.

The number of distant resonant objects beyond the 2:1 listed in the MPC from other surveys is $\sim103$, contrasting with 68 in the ensemble, of which 48 are from OSSOS.
Table~\ref{tab:totals} highlights that OSSOS provides the first TNOs found in a number of distant resonances. 
These include occupancy in the $a \sim 82$~au 9:2 resonance and the most distant securely resonant TNOs yet seen, in the $a \sim 130$~au 9:1 resonance \citep{Volk:2017inprep}. 
OSSOS has also found additional TNOs in other rarely seen distant resonances: 7 in the 3:1 and 3 in the 4:1 resonances \citep{Chiang:2003hb,Alexandersen:2016ki}.
The largest TNO found by OSSOS, the $H_r = 3.6 \pm 0.1$ dwarf planet candidate \texttt{o5s68} (2015 RR$_{245}$), is in the 9:2 resonance \citep{Bannister:2016AJ_OSSOSIV}; its long-term behaviour suggests that a large metastable population are cycling between the scattering population and high-order resonances.

\subsection{Extreme TNOs}
\label{sec:extreme}

\begin{figure}
\plotone{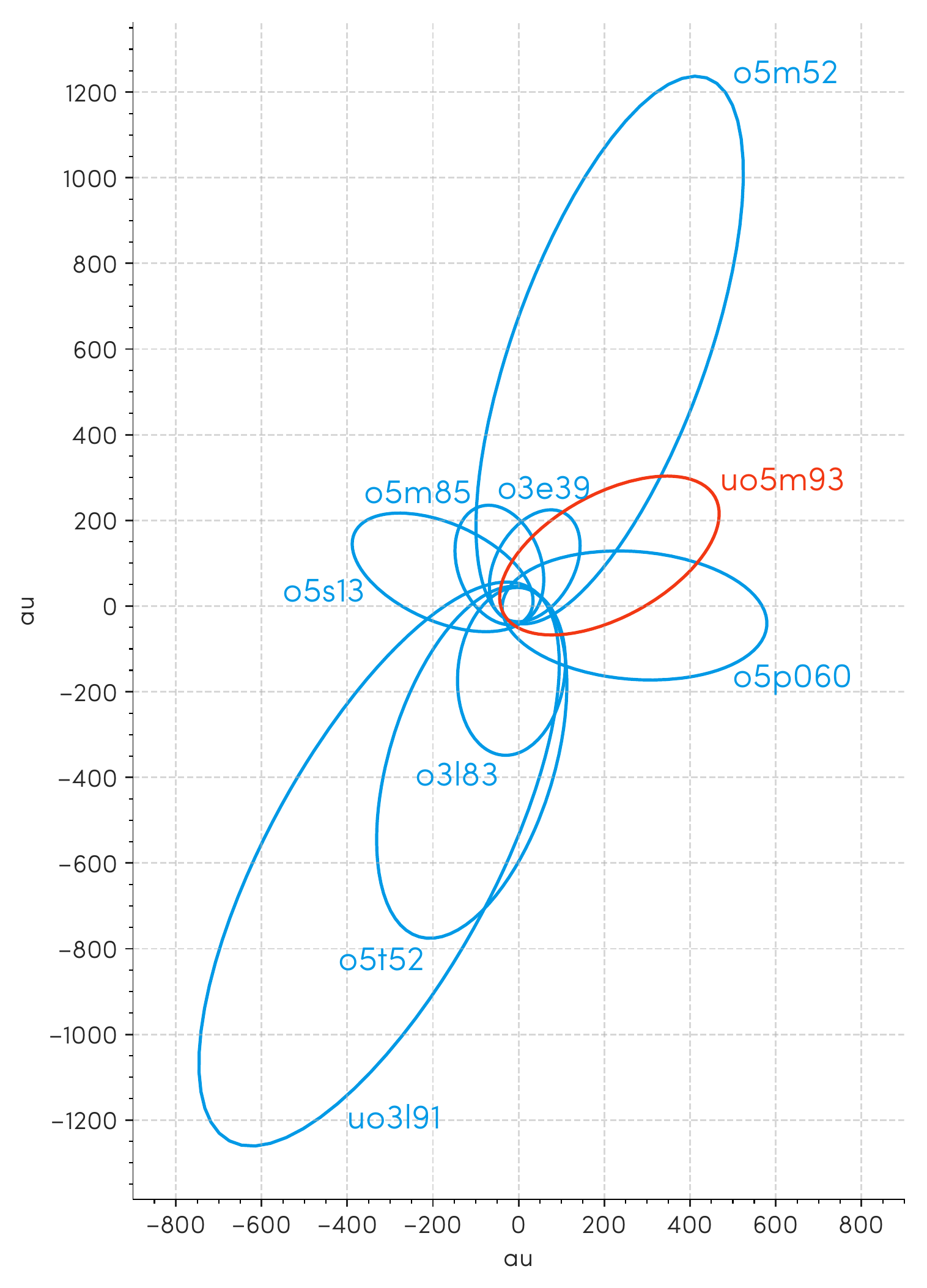}
\epsscale{\textwidth}
\caption{Ecliptic plane projection of the orbits of the nine $q > 30$, $a>150$~au TNOs discovered by OSSOS. 
The biases affecting discovery of the distant TNO population in OSSOS are quantified in \citet{Shankman:2017}, which lists the TNOs with orbits in blue. Newly reported in this work is {\tt uo5m93} (orbit in red), in Table~\ref{tab:discoveries}: catalogue of uncharacterized OSSOS discoveries.
These nine TNOs are consistent with being detected from a distribution of orbits that are intrinsically uniform in the angles $\omega$ and $\Omega$.
\label{fig:largea}}
\end{figure}

The OSSOS discoveries include nine TNOs with $q > 30$ au and $a > 150$ au, eight of which have $q > 38$ au. 
Their orbits are shown in Fig.~\ref{fig:largea}. 
Pronounced biases affect the discovery of minor planets on such distant orbits; the sensitivity of OSSOS to this population is quantified in \citet{Shankman:2017}.
There has been recent interest in the apparent angular clustering of the MPC-listed TNOs with $a > 150$ au orbits, which some have hypothesized as evidence for a massive distant planet \citep{Trujillo:2014ih,Batygin:2016ef}.
The sensitivity of OSSOS was applied to a distribution of simulated orbits that are uniform in their angles $\omega$, $\Omega$, and $\varpi$, to test if the detected OSSOS sample is consistent with being drawn from a uniform distribution.
In an analysis of the orbit distribution of the first eight of these TNOs, \citet{Shankman:2017} found that the OSSOS sample is consistent with being detected from that uniform distribution. 

We report a new minor planet in the $q > 30$~au, $a > 250$~au population.
The TNO \texttt{uo5m93}, an object in the uncharacterized OSSOS sample (Table~\ref{tab:discoveries}), was eventually tracked to a well-sampled four-opposition arc and found to have an orbit with $q=39.5$~au and $a=283$~au (Fig.~\ref{fig:largea}). 
Thorough inspection of the complete detection sample has minimised the possibility that any further $q > 30$~au, $a > 150$~au orbits remain concealed among the OSSOS {\tt nt} discoveries: fewer than 10 (with $d > 30$~au) of the 111 uncharacterized objects were untrackable (see Table~\ref{tab:discoveries}: uncharacterized, \texttt{nt} designations).
The spatial orientation of \texttt{uo5m93}'s orbit lies in an angular direction that is distinct from the hypothesised clustering, further weakening evidence for a lurking presence of intrinsic angular clustering.

Although this object is from the uncharacterized part of the sample, where our completeness limits are less well known, we re-performed the full statistical analysis of \citet{Shankman:2017}; the results are unchanged, and the OSSOS $q > 30$~au, $a > 150$~au nine-TNO sample is still consistent with being detected from a distribution of orbits that are intrinsically uniform in the angular elements.

\begin{figure*}
\plotone{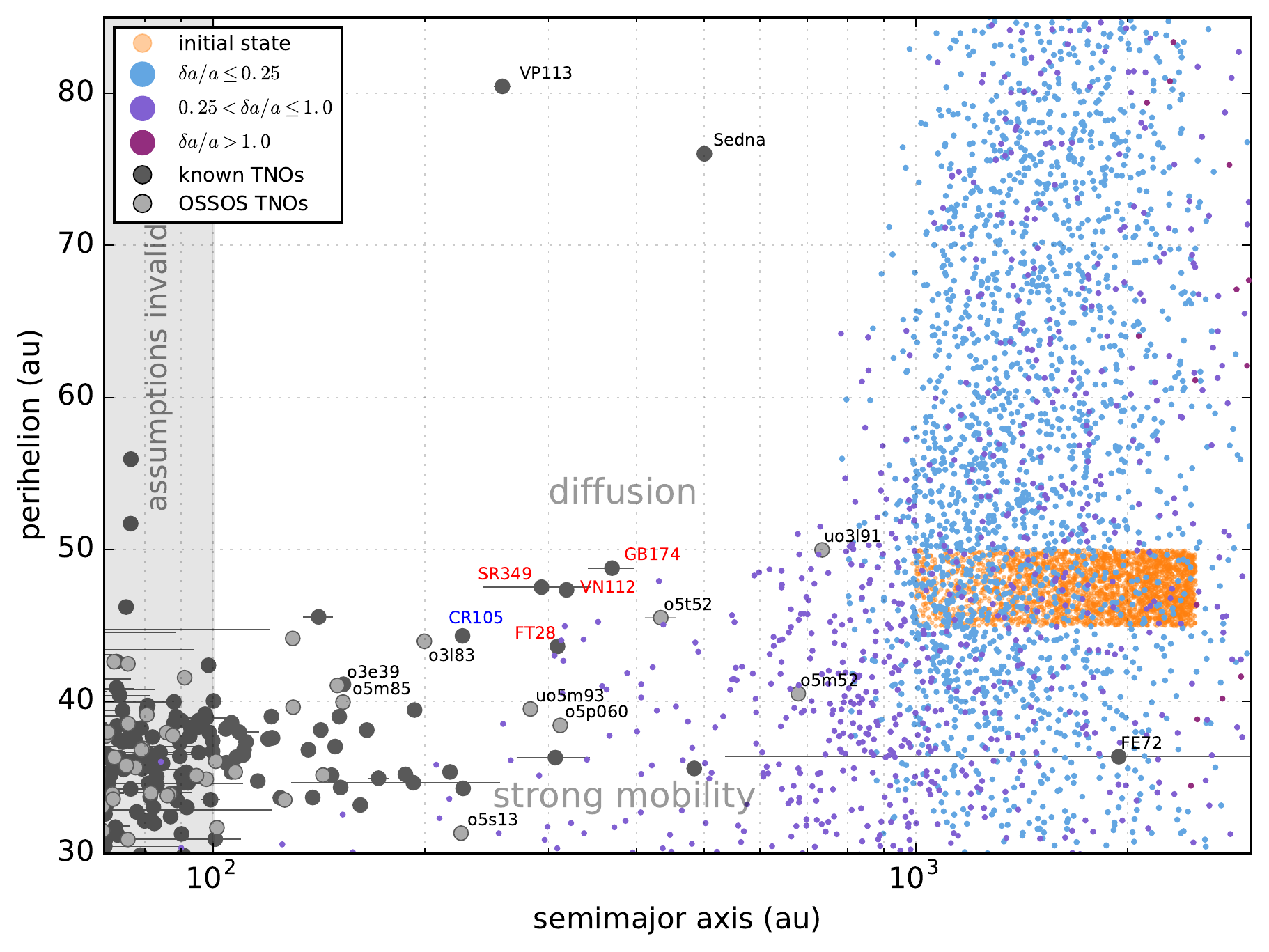}
\caption{Semi-major axis evolution demonstrating diffusion in large-$a$ TNOs with $q \lesssim 50$~au, updated from Fig. 5 in \citet{Bannister:2017arXivOSSOSV}. Known TNOs are shown with OSSOS discoveries (light grey) and the multi-opposition MPC TNOs (dark grey), together with the outcome of a model simulating a Gyr of evolution of particles started on orbits in the inner Oort cloud (start: orange; outcome: blue through purple). The nine $q > 30$, $a>150$~au OSSOS TNOs are named, as are the relevant MPC TNOs; 2000 CR$_{105}$ (name shown in blue) is in a high-order resonance, while the four red-labelled MPC TNOs exhibit diffusion in the integrations in \citet{Bannister:2017arXivOSSOSV}.
\label{fig:diffusion}}
\end{figure*}

Formation mechanisms for this distant population are not yet clear, and it remains an area of active investigation \citep[e.g.][]{Lawler:2016fg,Nesvorny:2017ku}. 
However, all the extreme TNO discoveries of OSSOS are consistent with a formation by random diffusion in semi-major axis, due to weak kicks at perihelion by Neptune, from orbits with semimajor axes in the inner fringe of the Oort cloud, as proposed in \citet{Bannister:2017arXivOSSOSV}. 
This is shown in Fig.~\ref{fig:diffusion}, which updates Fig. 5 in \citet{Bannister:2017arXivOSSOSV}: the OSSOS discoveries are in the phase spaces that can be populated by the combined mechanisms of diffusion, scattering and capture into weak distant resonances. 
We refer the reader to \citet{Bannister:2017arXivOSSOSV} for detailed discussion.  

\section{Summary}

The Outer Solar System Origins Survey acquired 8 TB of wide-field images of $\sim 170$ deg$^2$ of sky in 2013--2017 with CFHT's MegaCam.
These publicly-available data are calibrated to Data Release 1 from Gaia and Pan-STARRS.
Our analysis of the images has provided more than 37,000 astrometric measurements on the orbital arcs of 949 detections at heliocentric distances between 6 and 83 au; all points have been individually inspected for quality control.
From these detections, we present 840 discoveries of minor planets with precise quantification of their observational biases, ideal for use in testing models of TNO populations.
The orbital quality of 97\% of the 840 characterized discoveries is $\sigma_a < 0.1\%$.
This provides a significant improvement in both number per population, and orbital quality, over the sample in the MPC. 

We highlight several key results thus far from the OSSOS discovery sample and analysis papers:
\begin{itemize}
\item The scattering disk has an intrinsic population of $9 \times 10^4$ for $H_{r} < 8.66$ ($D > 100$~km), and the Centaur population is two orders of magnitude smaller.
\item The inner Kuiper belt has $a \gtrsim 37$~au, and its inclinations are consistent with those in the main belt.
\item Perihelia are consistently $q > 35$~au throughout the non-resonant Kuiper belt.
\item The low-inclination cold classicals have a ``kernel'' of population concentration in a tight parameter space at $a \sim 42.5-44.5$~au.
\item The low-inclination cold classical belt extends beyond the 2:1 resonance with Neptune to at least 49~au.
\item TNOs in the 2:1 resonance have a colder inclination distribution than those in the closer 3:2 resonance.
\item The distant $a > 50$~au resonances are more heavily populated than current Neptune migration models predict.
\item Securely occupied resonances exist out to at least $a \sim 130$~au. Their long-term behaviour on timescales of more than a hundred Myr suggests a large metastable population are cycling between the scattering disk and high-order resonances.
\item Our nine $q > 30$, $a> 150$~au TNOs are consistent with being detected from an orbit distribution intrinsically uniform in the angles $\omega$ and $\Omega$, and with formation via the mechanisms of diffusion, scattering, and capture into distant resonances.
\end{itemize}

In contrast, the MPC sample is from a variety of surveys, largely without bias characterization.
Thus, the OSSOS minor planets form a unique set: a sample that is half as large as the entire MPC inventory, yet with perfect tracking efficiency for trans-Neptunian orbits, and quantifiable biases.
We provide a survey simulator that includes the specifications of OSSOS and three other smaller surveys with well-quantified biases, for a total discovery sample of 1142 bias-characterized TNOs with which to test population models. 
We look forward to seeing the community use this powerful tool for diagnosing the inventory and history of our Solar System.

\acknowledgments

The authors acknowledge the sacred nature of Maunakea, and appreciate the opportunity to observe from the mountain.
This project could not have been a success without the staff of the Canada-France-Hawaii telescope: thank you all for your amazing dedication, enthusiasm, and your concerted effort throughout our five years.
CFHT is operated by the National Research Council (NRC) of Canada, the Institute National des Sciences de l'Universe of the Centre National de la Recherche Scientifique (CNRS) of France, and the University of Hawaii, with OSSOS receiving additional access due to contributions from the Institute of Astronomy and Astrophysics, Academia Sinica, Taiwan.
Data were produced and hosted at the Canadian Astronomy Data Centre; processing and analysis were performed using computing and storage capacity provided by the Canadian Advanced Network For Astronomy Research (CANFAR).

M.T.B. appreciates support during OSSOS from UK STFC grant ST/L000709/1, the National Research Council of Canada, and the National Science and Engineering Research Council of Canada.
K.V. acknowledges support from NASA grants NNX14AG93G and NNX15AH59G.
R.I.D. acknowledges the Center for Exoplanets and Habitable Worlds, which is supported by the Pennsylvania State University, the Eberly College of Science, and the Pennsylvania Space Grant Consortium.
M.J. acknowledges support from the Slovak Grant Agency for Science (grant VEGA No. 2/0037/18).
S.M.L. gratefully acknowledges support from the NRC-Canada Plaskett Fellowship.
M.E.S. was supported by Gemini Observatory, which is operated by the Association of Universities for Research in Astronomy, Inc., on behalf of the international Gemini partnership of Argentina, Brazil, Canada, Chile, and the United States of America. MES was also supported in part by an Academia Sinica Postdoctoral Fellowship.

This research has made use of NASA's Astrophysics Data System Bibliographic Services, the JPL HORIZONS web interface (\url{https://ssd.jpl.nasa.gov/horizons.cgi}), and data and services provided by the International Astronomical Union's Minor Planet Center. 

\facility{CFHT (MegaPrime)}
\software{
Astropy \citep{TheAstropyCollaboration:2013cd}; 
matplotlib \citep{Hunter:2007}, 
scipy \citep{jones:2001}, 
GNU parallel \citep{Tange2011a},  
SExtractor \citep{Bertin:1996}, 
IRAF \citep{Tody:1986},
DAOphot \citep{Stetson:1987},
WCStools, 
SuperMongo.
}

\bibliographystyle{aasjournal}
\bibliography{references}

\end{document}